\documentclass[11pt]{article}
\usepackage{graphicx} 
\usepackage{amsmath}
\usepackage{amssymb}
\usepackage{caption}
\usepackage{overpic}
\usepackage{hyperref}
\usepackage{comment}
\usepackage{dcolumn}
\usepackage{bm}
\usepackage[margin=1in]{geometry}
\usepackage[utf8]{inputenc}
\usepackage[T1]{fontenc}
\usepackage{mathptmx}
\usepackage{etoolbox}
\usepackage{siunitx}
\usepackage{subcaption}
\usepackage{pgfplotstable}
\usepackage{float}
\usepackage{amssymb}
\usepackage[numbers,sort&compress]{natbib}
\usepackage{placeins}
\newtheorem{remark}{Remark}

\newcommand{\anna}[2][cyan]{\textcolor{#1}{#2}}

\title{Enhancing the accuracy of under-resolved numerical simulations of 
atmospheric flows with 
super resolution 
}

\begin{document}

\author{Armin Sheidani$^{1 ,2}$, Michele Girfoglio$^3$, Annalisa Quaini$^{4}$, Gianluigi Rozza$^1$}

\maketitle

\begin{center}
\footnotesize
$^1$ mathLab, Mathematics Area, SISSA, via Bonomea, 265, Trieste, I-34136, Italy \\ $^2$ Fluids and Flows, Department of Applied Physics and Science Education, Eindhoven University of Technology, Eindhoven, P.O. Box 513, 5600 MB, The Netherlands \\ $^3$ University of Palermo, Faculty of Engineering, Viale delle Scienze, Ed. 7, 90128 Palermo, Italy \\ $^4$ Department of Mathematics, University of Houston, 3551 Cullen Blvd, Houston TX 77204, USA

\end{center}

\begin{abstract}
Super-resolution (SR) techniques based on deep learning have recently emerged as a promising approach to enhance the spatial resolution of computational fluid dynamics simulations while containing computational cost. In this paper, we investigate several SR architectures to improve coarse-grid simulations of mesoscale atmospheric flows, with training data generated from simulations of the weakly compressible Euler equations. We compare a baseline convolutional neural network (CNN), an attention-enhanced CNN, a multi-scale CNN designed to capture flow structures across different spatial scales, and a diffusion-based SR model. The methods are evaluated on two standard atmospheric benchmarks: the rising thermal bubble and the density current. Results show that the baseline CNN can accurately reconstruct simpler flow features, while more complex flows require multi-scale architectures. Overall, SR based on the multi-scale CNN provides the best balance of accuracy, robustness, and computational efficiency, outperforming even a state-of-the-art diffusion-based approach. We also analyze the sensitivity of the models to the size of the training dataset, highlighting limitations and trade-offs of the proposed SR strategies.
\end{abstract}

\section{Introduction}\label{sec:intro}

Super-resolution (SR) is a set of methods used to reconstruct a high-resolution signal or image from one or more low-resolution observations.
See, e.g., \cite{park2003super,farsiu2004advances}.
In simple terms, SR tries to increase the spatial resolution beyond what is present in the original data.
It finds applications in various fields, such as medical imaging \cite{li2021review, greenspan2009super, isaac2015super,el2024single, huang2024deep}, satellite imaging \cite{constantinou2024leveraging, karwowska2022using, he2021spatial, chen2020monitoring}, and digital photography \cite{pashaei2020deep, bhat2021deep, ma2020quanta, hansen2021super}. 
From a computational perspective, SR is
an ill-posed inverse problem that often needs regularization or learned priors. 
Convolutional neural networks (CNN)
can learn strong priors from data and thus have become a very popular class of methods for SR. In fact, by employing large datasets of paired low- and high-resolution samples, CNNs can learn the underlying structures in the data, allowing for accurate and efficient reconstruction of fine-scale details.

Recently, SR methods have found application in computational fluid dynamics (CFD) with the goal of enhancing the spatial 
resolution of fluid flow simulations. See, e.g., \cite{sofos2025review,zhang2025temporal,page2025super,cheng2025improved, hu2024super, trinh20243d, saitta2024implicit, pang2024deep, li2022using}.
Through SR, high-resolution details can be extracted from low-resolution flow
simulations to achieve higher accuracy while containing the computational cost.
This is particularly appealing for convection-dominated flows, whose 
vortices and eddies span an increasingly large range of scales as the inertial forces become more and more dominant over the viscous forces.
For this kind of flows, accurate simulations based on traditional computational methods, e.g., Raynolds Average Navier-Stokes (RANS) and Large Eddy Simulations (LES), carry a high computational cost. 
The goal to reduce the computational cost of these methods while maintaining a reasonable level of accuracy
has motivated a vast body of literature
from the CFD community.
This is one reason why CNN-based SR is gaining popularity for CFD applications.
Standard SR techniques and multi-scale versions \cite{bi2022flowsrnet}
have been used to 
reconstruct test cases, such as the flow past a cylinder in turbulent regime \cite{fukami2019super}, and more complex flows, like airflows in data centers \cite{hu2024super}. SR methods have also been coupled with other techniques, such as feature recognition \cite{tang2024super}
and transformers \cite{xu2023super},
to reduce the sensitivity to the low-resolution input. 
For a survey of SR for fluid flows, we refer to \cite{fukami2023super}.

In this paper, we focus on 
mesoscale atmospheric flow, which is heavily convection-dominated. While the development
of accurate RANS (see, e.g., \cite{van2020rossby,cindori2022comparison, van2021inflow, bellegoni2023extended, han2020rans})
and LES (see, e.g., \cite{stoll2020large,zahn2024setting, cotteleer2024flow,liu2023wavelet,clinco2023filter})
methods for atmospheric flows is still ongoing, the CFD community has increasingly turned to machine learning 
techniques. These include neural networks (e.g., \cite{yuval2023neural}),
convolutional autoencodencoders (e.g., \cite{zhang2023toward}), 
physics-informed neural network (e.g., \cite{angriman2023assimilation}), 
CNN and time series prediction using Long Short Term Memory (LSTM) and transformer (e.g., \cite{cui2023deep}).
CNN-based approaches have been used 
both for SR~\citep{yasuda2025two, yasuda2022super} and statistical downscaling \cite{jha2025deep,getter2024statistical,hohlein2020comparative}, which finds correlations between large-scale atmospheric patterns from global climate model and local climate data to produce high-resolution, local-scale climate information. While these preliminary works show promising results, several questions remain open. For example, little attention has been devoted to the understanding 
which CNN architecture is best suited for given flow field characteristics.
In this work, we address this by systematically comparing several deep learning architectures for SR. 
It also remains unclear to what extent a neural network trained on high-fidelity data can recover flow structures that are not explicitly resolved in low-resolution simulations. 
Thus, we investigate the ability of the different SR models to reconstruct fine-scale structures from coarse simulations and assess the influence of training dataset size on reconstruction accuracy.

To generate training data for different SR approaches, we consider two models. We call one of them artificial viscosity (AV) because it introduces a constant artificial viscosity to the weakly compressible Euler equations as a crude LES method
(see, e.g., \cite{yu2015localized,GQR_OF_clima,clinco2023filter}), while the other involves a standard Smagorinsky closure. Low-resolution images are generated from coarse-mesh simulations with the AV model, while the high-resolution images that serve as training targets come from fine mesh simulations using either the AV or
the Smagorinsky model. The reason behind the choice of these two models is that one (Smagornsky) reveals 
more flow structures than the other (AV).  
Several deep-learning architectures are explored to learn the low-resolution to high-resolution mapping: a baseline CNN composed of repeated blocks of bilinear upsampling, convolutional layers, and nonlinear activation functions; an attention-enhanced CNN that introduces an attention block to allow the network to focus on the most relevant spatial features; and a multi-scale CNN with parallel convolutional branches using filters of different sizes to capture
flow structures at multiple spatial scales. 
We train these networks by minimizing the mean squared error between low-resolution and reference high-resolution fields using the Adam optimizer. 
In addition, we evaluate a diffusion-based SR model, which
is trained to reconstruct the high-resolution field from a low-resolution input through a gradual denoising process. During training, noise is progressively added to high-resolution images, and a neural network, based on a U-Net architecture, learns to predict and remove this noise while being guided by the corresponding upsampled low-resolution image. 
At inference time, the process starts from random noise and iteratively removes the noise while conditioning on the low-resolution input, gradually producing a high-resolution reconstruction that is consistent with the coarse data. 
The various SR methods are tested on two standard atmospheric benchmarks to assess their ability to enhance coarse-grid simulations.

The rest of the paper is organized as follows. In Sec.~\ref{sec:pbd},
we state the compressible Euler equations for low Mach, stratified flows, which model dry atmospheric flows, and discuss their
time and space discretization. Sec.\ref{sec:SR} briefly covers the foundational mathematical principles of super resolution, including a detailed explanation of the methodologies employed within this study. 
Numerical results are reported in Sec.~\ref{Sec:res}. Conclusions are drawn in Sec.~\ref{sec:concl}.

\section{Problem definition}
\label{sec:pbd}


We consider the dynamics of the dry atmosphere
over a time interval of interest $(0,t_f]$
in a spatial domain $\Omega$, disregarding the effects of moisture, solar radiation, and ground heat flux. 
Such dynamics are governed by the weakly compressible Euler equations, with the 
assumption that dry air behaves as an ideal gas. In order to write these equations, we need to introduce some notation.
Let $\rho$ represent air density, $\mathbf{u} = (u, v, w)$ the wind velocity, and $e$ the total energy density.
Note that we have  $e = c_v T + |\mathbf{u}|^2/2 + g z$ , where $c_v$ is the specific heat capacity at constant volume, $T$ is the absolute temperature, $g$ is the gravitational constant, and $z$ is the vertical coordinate. Then, the
weakly compressible Euler equations can be written as: 
\begin{align}
&\frac{\partial \rho}{\partial t} + \nabla \cdot (\rho \mathbf{u}) = 0 &&\text{in } \Omega \times (0,t_f], \label{eq:mass}  \\
&\frac{\partial (\rho \mathbf{u})}{\partial t} +  \nabla \cdot (\rho \mathbf{u} \otimes \mathbf{u}) + \nabla p   + \rho g \widehat{\mathbf{k}} = \boldsymbol{0} &&\text{in } \Omega \times (0,t_f],  \label{eq:mom} \\
&\frac{\partial (\rho e)}{\partial t} +  \nabla \cdot (\rho \mathbf{u} e) + \nabla \cdot (p \mathbf{u}) = 0 &&\text{in } \Omega \times (0,t_f],
\label{eq:ent}
\end{align}
where $\widehat{\mathbf{k}}$ is the unit vector aligned with the vertical axis $z$ and $p$ is pressure. Eq.~\eqref{eq:mass}-\eqref{eq:ent}
state conservation of mass, momentum, and total energy. 
To close system \eqref{eq:mass}-\eqref{eq:ent}, we need a thermodynamic equation of state for $p$. Based on the assumption that dry air behaves as an ideal gas, we have: 
\begin{align}
p = \rho R T, 
\label{eq:p}
\end{align}
where $R$ is the specific gas constant of dry air. 

We express the pressure as the sum of a fluctuation $p'$ relative to a background state: 
\begin{align}
p = p_g + \rho g z + p', \label{eq:p_split}
\end{align}
where $p_g$ = $10^5$ Pa is the atmospheric pressure at the ground. 
By plugging \eqref{eq:p_split} into eq.~\eqref{eq:mom}, we obtain:
\begin{align}
\frac{\partial (\rho \mathbf{u})}{\partial t} +  \nabla \cdot (\rho \mathbf{u} \otimes \mathbf{u}) + \nabla p' + gz \nabla \rho =0\quad \text{in } \Omega \times (0,t_f].  \label{eq:mom_split}
\end{align}

\begin{remark}
Pressure fluctuation $p'$ in \eqref{eq:p_split} is different from 0 also in hydrostatic balance conditions because 
$\rho$ does not remain constant in the compressible regime. This is due to the definition of the background state in \eqref{eq:p_split}, which has been widely 
used in the literature (see, e.g., 
 \cite{Zancanaro, Weller1998, jasakphd} and references therein).
Other possible definitions of the pressure fluctuation
are discussed and compared 
in, e.g., \cite{giraldo_2008,GIRFOGLIO2025106510}.
The strategy presented in this paper works regardless of the definition of $p'$.
\end{remark}

Let $c_{p}$ be the specific heat capacity at constant pressure for dry air. Moreover, let $K$
be the kinetic energy density and $h$ the specific enthalpy: 
\begin{align}
K = |\mathbf{u}|^2/2, \quad  h = c_v T + p/\rho = c_p T, \label{eq:K}
\end{align}
respectively. The total energy density $e$ can be expressed as $e = h - p/\rho + K + gz$ and  eq.~\eqref{eq:ent} becomes:
\begin{align}
\frac{\partial (\rho h)}{\partial t} +  \nabla \cdot (\rho \mathbf{u} h) + 
\frac{\partial (\rho K)}{\partial t} +  \nabla \cdot (\rho \mathbf{u} K) - \dfrac{\partial p}{\partial t}  +  
\rho g \mathbf{u} \cdot \widehat{\mathbf{k}} = 0,
\label{eq:over_ent}
\end{align}
where we have used eq. \eqref{eq:mass} for further simplification. 




Equations \eqref{eq:mass}, \eqref{eq:p}-\eqref{eq:mom_split}, and \eqref{eq:over_ent}
do not have any built-in mechanism to dissipate
energy. Hence, numerical methods for this problem 
typically add artificial diffusion for stability.
One way to to this it through a 
a Large Eddy Simulation (LES) approach, which amounts to adding an artificial diffusion
term to eqs. \eqref{eq:mom_split} and \eqref{eq:over_ent}:
\begin{align}
&\frac{\partial (\rho \mathbf{u})}{\partial t} +  \nabla \cdot (\rho \mathbf{u} \otimes \mathbf{u}) + \nabla p' + gz \nabla \rho -  \nabla \cdot (2 \mu_a \boldsymbol{\epsilon}(\mathbf{u})) + \nabla \left(\frac{2}{3}\mu_a \nabla \cdot \mathbf{u} \right)= 0 &&\text{in } \Omega \times (0,t_f],  \label{eq:mom_LES} \\
&\frac{\partial (\rho h)}{\partial t} +  \nabla \cdot (\rho \mathbf{u} h) + 
\frac{\partial (\rho K)}{\partial t} +  \nabla \cdot (\rho \mathbf{u} K) - \dfrac{\partial p}{\partial t}  +  
\rho g \mathbf{u} \cdot \widehat{\mathbf{k}}  - \nabla \cdot \left(\frac{\mu_a}{Pr} \nabla h \right) = 0 &&\text{in } \Omega \times (0,t_f],
\label{eq:ent_LES}
\end{align}
where $\boldsymbol{\epsilon}(\mathbf{u}) = (\nabla \mathbf{u} + (\nabla \mathbf{u})^T)/2$ denotes the strain-rate tensor and $Pr$ stands for the Prandtl number, which is the dimensionless ratio of momentum diffusivity to thermal diffusivity.
The key parameter in \eqref{eq:mom_LES}-\eqref{eq:ent_LES} is $\mu_a$, 
which represents an artificial viscosity and has different definitions
depending on the different LES models.

In this work, we consider two strategies to set $\mu_a$. The first strategy is to set a constant ad hoc value taken from the literature. Hereinafter, we refer to this model as AV, where AV stands for
artificial viscosity. We note that this is a very crude LES model since $\mu_a$ remains constant
in space and time. 
The second strategy is
the classical Smagorinsky model \cite{smagorinsky1963},
for which the artificial viscosity is given by:
\begin{align}
\mu_a = \rho (C_s \delta)^2 \sqrt{ 2 \boldsymbol{\epsilon} : \boldsymbol{\epsilon}}, \quad C_s^2 = C_k \sqrt{\dfrac{C_k}{C_{\epsilon}}}, \label{eq:smago}
\end{align}
where $\delta$ is the so-called filter width, and $C_k$ and $C_\epsilon$ are model parameters.
The values of these parameters are not universal and need to be properly selected based on the problem at hand. 
In Sec.~\ref{Sec:res}, we will specify
the values of these parameters for each
benchmarks under consideration. 

Finally,
a quantity of interest in atmospheric studies is the potential temperature, defined as
\begin{align}
\theta = \frac{T}{\pi}, \quad \pi = \left( \frac{p}{p_g} \right)^{\frac{R}{c_{p}}}, \quad 
p_g = 10^5 \text{ Pa}. \label{eq:theta}
\end{align}
The potential temperature represents the temperature a parcel of dry air would attain if adiabatically expanded or compressed to reference pressure $p_g$, 
which is the atmospheric pressure at ground level. 
Similarly to what done for the pressure, we introduce
the potential temperature fluctuation $\theta'$ 
with respect to a background state $\theta_0$, which we take to be an average hydrostatic value:
\begin{align}
\theta'(x,y,z,t) = 
\theta(x,y,z,t) - \theta_0(z) . \label{eq:theta_split} 
\end{align}
Note that $\theta_0$ depends only on the vertical coordinate $z$

\subsection{Time and space discretization}\label{sec:st_disc}

In this section, we are going to briefly introduce the space and time discretization for the model
\eqref{eq:mass}, \eqref{eq:p}-\eqref{eq:p_split}, \eqref{eq:mom_LES}-\eqref{eq:ent_LES} and the solution algorithm we adopt to compute the approximated solution. The reader interested in more details is referred to
\cite{GQR_OF_clima,GirfoglioFVCA10}.

For the space discretization,
we use a Finite Volume method. This requires to partition the computational domain $\Omega$ into cells or control volumes $\Omega_i$, with $i = 1, \dots, N_{c}$, where $N_{c}$ is the total number of cells in the mesh. Let  \textbf{A}$_j$ be the surface vector of each face of the control volume (i.e., the surface area of the face multiplied by the outward normal), 
with $j = 1, \dots, M$. With the subindex $i$, we will indicate a variable average in control volume $\Omega_i$. For time discretization, we adopt the backward Euler scheme. We partition the time interval
$(0, t_f]$ with a time step $\Delta t$ and approximate the solution
at time $t^n = n \Delta t$, with $n = 0, ..., N_f$ and $t_f = N_f \Delta t$. We will denote by $y^n$ the approximation of a generic quantity $y$ at the time $t^n$. 

For computational efficiency, once the problem is discretized in space and time we adopt a three-step splitting approach, which reads as follows: given $\rho^0$, $\mathbf{u}^0$, $h^0$,  $p^0$, and $T^0$, set $K^0 = |\mathbf{u}^0|^2/2$ and for $n \geq 0$ find solution $(\rho^{n+1}_i, \mathbf{u}^{n+1}_i,h^{n+1}_i,K^{n+1}_i, p^{n+1}_i, p'^{,n+1}_i, T^{n+1}_i)$ by performing
the following three steps:

\begin{itemize}
\item[-] \emph{Step 1}: find the first intermediate density ${\rho}_i^{n+\frac{1}{3}}$, intermediate velocity ${\mathbf{u}}_i^{n+\frac{1}{3}}$ and associated kinetic energy density $K_i^{n+\frac{1}{3}}$ such that
\begin{align}
    \frac{\rho^{n+\frac{1}{3}}}{\Delta t} &= b_{\rho}^{n+1} - \nabla \cdot (\rho^n \mathbf{u}^n), \label{eq:step1_sd} \\[10pt]
    \frac{\rho^{n+\frac{1}{3}} \mathbf{u}^{n+\frac{1}{3}}}{\Delta t} &+ \nabla \cdot \left( \rho^n \mathbf{u}^n \otimes \mathbf{u}^{n+\frac{1}{3}} \right) + \nabla p^n \nonumber \\
    &- \nabla \cdot \left( 2 \mu_a^n \boldsymbol{\epsilon} \left( \mathbf{u}^{n+\frac{1}{3}} \right) \right) + \nabla \left( \frac{2}{3} \mu_a^n \nabla \cdot \mathbf{u}^n \right) = \mathbf{b}_{\mathbf{u}}^{n+1}, \label{eq:step1_2sd} \\[10pt]
    K^{n+\frac{1}{3}} &= \frac{|\mathbf{u}^{n+\frac{1}{3}}|^2}{2}. \label{eq:step1_3sd}
\end{align}
Note that Eq.~\ref{eq:step1_3sd} is a predictor step, and the associated corrector step is Eq.~\ref{eq:step3_4sd}.
\item[-] \emph{Step 2}: find average specific enthalpy $h^{n+1}_i$, temperature $T^{n+1}_i$, and the second intermediate density ${\rho}^{n+\frac{2}{3}}_i$ in control volume $\Omega_i$ such that
\begin{align}
    \frac{\rho^{n+\frac{1}{3}} h^{n+1}}{\Delta t} &+ \nabla \cdot (\rho^n \mathbf{u}^n h^{n+1}) - \nabla \cdot \left( \frac{\mu_a^n}{\text{Pr}} \nabla h^{n+1} \right) \label{eq:step2_sd} \\
    &= \tilde{b}_e^n - \frac{\rho^{n+\frac{1}{3}} K^{n+\frac{1}{3}}}{\Delta t} - \nabla \cdot \left( \rho^n \mathbf{u}^n K^{n+\frac{1}{3}} \right) + \frac{p^n}{\Delta t} \nonumber  \\
    &- \rho^{n+\frac{1}{3}} g \mathbf{u}^{n+\frac{1}{3}} \cdot \mathbf{\hat{k}}, \nonumber \\[10pt]
    h^{n+1} - c_p T^{n+1} &= h^n - c_p T^n, \label{eq:step2_2sd_2} \\[10pt]
    \rho^{n+\frac{2}{3}} R T^{n+1} &= p^n, \label{eq:step2_2sd}
\end{align}
where ${b}_e^{n} = (\rho^n h^n + \rho^n K^{n-1} - p^{n-1}) / \Delta t$.
Eq.~\eqref{eq:step2_2sd_2} will be used to compute the end-of-step temperature, while eq.~\eqref{eq:step2_2sd} is a predictor step for density. The associated correction will come from eq.~\eqref{eq:step3_2sd_2}.
\item[-] \emph{Step 3}: find end-of-step velocity $\mathbf{u}^{n+1}$ and associated kinetic energy density $K^{n+1}$, pressure $p^{n+1}$ and pressure fluctuation $p'^{,n+1}$, and end-of-step density $\rho^{n+1}$ 
such that 
\begin{align}
    \frac{\rho^{n+\frac{1}{3}} \mathbf{u}^{n+1}}{\Delta t} &+ \nabla \cdot (\rho^n \mathbf{u}^n \otimes \mathbf{u}^{n+1}) + \nabla p^{n+1} - \nabla \cdot (2\mu_a^n \boldsymbol{\epsilon}(\mathbf{u}^{n+1})) \nonumber \\
    &+ \nabla \left( \frac{2}{3} \mu_a^n \nabla \cdot \mathbf{u}^n \right) = \mathbf{b}_{\mathbf{u}}^{n+1}, \label{eq:step3_sd} \\[10pt]
    p^{n+1} - p^{\prime, n+1} &= \rho^{n+\frac{2}{3}} gz, \label{eq:step3_2sd} \\[10pt]
    p^{n+1} - \rho^{n+1} RT^{n+1} &= 0, \label{eq:step3_2sd_2} \\[10pt]
    \frac{\rho^{n+\frac{2}{3}}}{\Delta t} + \nabla \cdot \left( \rho^{n+\frac{2}{3}} \mathbf{u}^{n+1} \right) &= b_{\rho}^{n+1}, \label{eq:step3_3sd} \\[10pt]
    K^{n+1} &= \frac{|\mathbf{u}^{n+1}|^2}{2}. \label{eq:step3_4sd}
\end{align}
It should be noted that in Eq.~\ref{eq:step3_sd}, one diffusive term is maintained as implicit while the other is handled explicitly, a practice that is customary in OpenFOAM. In order to decouple the computation of these two variables, we use the PISO algorithm \cite{Weller1998}.
\end{itemize}

For the approximation of all the Laplacian, gradient and divergence terms in the equations above, we choose second-order accurate schemes.


\section{The super-resolution based on convolutional neural networks} \label{sec:SR}


SR provides an approximation of a high-resolution image ${I}_{HR}$ from a low-resolution image \( I_{LR} \). This is done though a map, called super-resolution function and denoted with $SR$, so that the 
approximation \( \hat{I}_{HR} \) of ${I}_{HR}$
is given by $\hat{I}_{HR} = SR(I_{LR})$.
This section describes several deep-learning architectures we will use to learn the super-resolution function.
The images are plots of 2D
solutions computed by the algorithm reported in Sec.~\ref{sec:st_disc}. Specifically, 
\( I_{LR} \) is a plot of a solution at a given time instance computed by 
the AV model with a very coarse mesh
and, as such, expected to be of poor quality,
while \( I_{HR} \) is a plot of the solution at the same time instance computed by either the AV model or by the Smagorinsky
model with a fine mesh. 

The plots of two-dimensional numerical solutions
consist of two-dimensional representations of pixels in rectangular coordinates. They can be represented as a tensor of dimensions $w \times h \times c$, where $w$ is the width of the image and $h$ is its height, i.e., the number of pixels along the horizontal direction and vertical direction, and $c$ is the number of channels. 
For a color image, $c$ corresponds to the number of colors that properly combined give each pixel of the image. Usually, we have $c = 3$, the three colors being red, green, and blue. In our case,  $c$ is related to the output of the numerical simulation for a given scalar variable and thus we set $c = 1$ because 
we view it as inherently monochromatic, providing a single, scalar value per spatial point. The computed scalar variable 
is visualized with a colormap, but the underlying data structure per pixel is one-dimensional.
We note that $w$, $h$, and $c$ will change as the image passes through the network.


The architectures we will consider to perform the super-resolution are described
in order of increasing complexity. We give an
acronym to each architecture that will be used in the rest of the paper.

\begin{figure}[htb!]
    \centering
    \begin{overpic}[width=0.95\textwidth]
{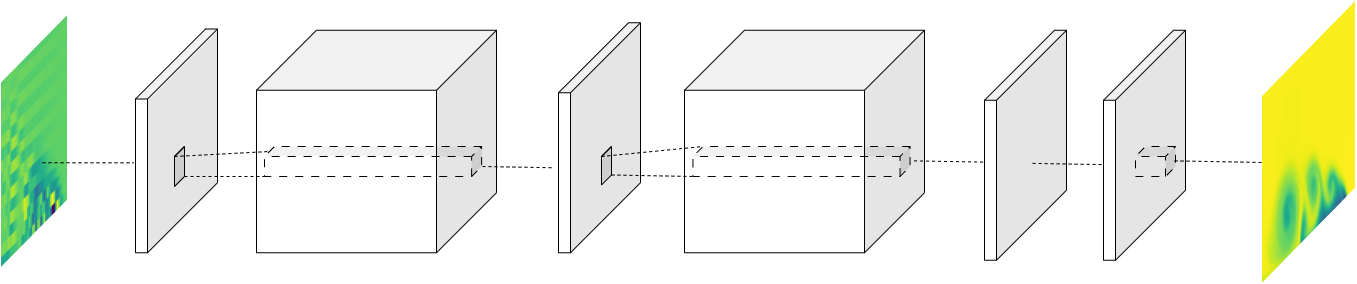}
\put(2,0){$I_{LR}$}
\put(11,0){U1}
\put(25,5){{$K$}}
\put(24,0){C1}
\put(37,10){AF}
\put(43,0){U2}
\put(56,0){C2}
\put(57,5){{$K$}}
\put(69,10){AF}
\put(75,0){A}
\put(85,0){C3}
\put(83.8,5.5){{$K$}}
\put(97,0){$\hat{I}_{HR}$}
\end{overpic}
    \caption{Schematic diagram of a \emph{A-CNN} architecture: U$\alpha$ 
    denotes upsampling layer $\alpha$, C$\beta$ denotes 
    convolution layer $\beta$, $K$ is a filter matrix, 
    AF represents an activation function,
    and A denotes an attention block. 
    }
    \label{fig:CNN_blocks}
\end{figure}
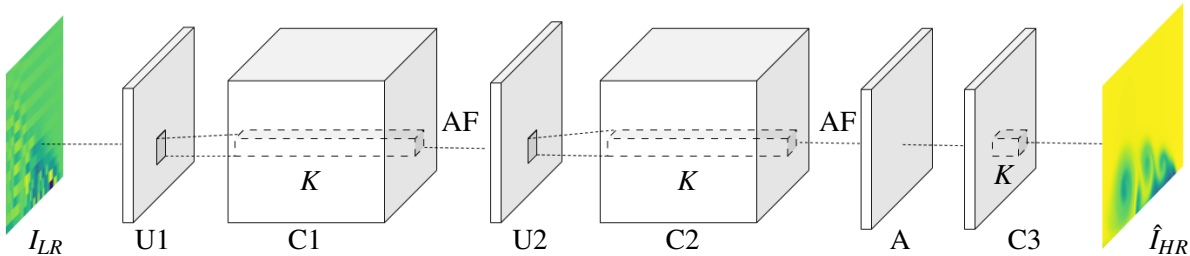

{\bf \emph{CNN}}. 
As mentioned in Sec.~\ref{sec:intro},
one of the most successful approaches for SR is based on deep learning techniques, particularly CNNs. 
Here, we introduce the concepts that are relevant to our work. The reader interested in a detailed discussion on the mathematics behind CNNs and its variants, 
is referred to, e.g., \citep{lecun1995convolutional, krizhevsky2017imagenet, he2016deep}.
The typical architecture used for SR applications consists of two main elements: \emph{upsampling layers} and \emph{convolutation layers}. 
An {upsampling layer} is applied to increase the spatial dimensions (i.e., $w$ and $h$) of the initial input image, i.e., 
$I_{LR}$, or an intermediate input image. 
This can be done with several techniques, e.g., interpolation methods, deconvolution, and pixel shuffle. See, e.g., \cite{lim2017enhanced, shi2016real, zhou2016learning, radford2015unsupervised} for more details. In this work, we adopt a bilinear interpolation method that, for every pixel, takes a weighted average of the four nearest pixel values to estimate the new pixel value, generating an output image of dimensions $w' \times h' \times c$, with $w' = 2w$ and $h' = 2h$. This $w' \times h' \times c$ image is then processed by a {convolutional layer}. 
The convolution operation involves a filter, also known as kernel,
that 
moves over the image systematically and
it can be expressed as: 
\begin{equation*}
I_{LR} * K = \sum_{m=0}^{M-1} \sum_{n=0}^{N-1} I_{LR}(i+m, j+n, l) \cdot K(m, n), \quad {i = 1, \dots, w',~j = 1, \dots, h', ~l= 1, \dots, c',}
\end{equation*}
where $K$ is the filter matrix of dimensions $M \times N$.
The output of this operation $I_{LR} * K$ is called feature map. 
Typically, multiple filters  are used in each convolutional layer to capture different types of features in the input data and the number of filters is set by trial and error. 
The feature map resulting from  each filter corresponds to a new specific channel. 
Hence, the number of channels (i.e., the depth of the convolution layer) changes from the initial value $c$
to a new value $c'$, 
increasing as the image goes deeper into the network.
After one upsampling and one convolutional layer, 
an activation function is applied to the output image
of size $w' \times h' \times c'$. 
This enables the network to learn complex patterns by introducing non-linearity into the model.  
See Fig.~\ref{fig:CNN_blocks}, where the low-resolution image $I_{LR}$ is processed by two blocks of upsampling layer + convolution layer + activation function (U1-C1-AF and U2-C2-AF) and then further elaborated as explained in the next paragraph.
The typical SR architecture, 
which we will simply denote with \emph{CNN}, features
several blocks of {upsampling layer + convolutional layer + activation function}, with the number of such blocks decided by trial and error.

We will train \emph{CNN} by minimizing the mean squared error (MSE) between the high-resolution image \( I_{LR} \) and the predicted high-resolution image \( \hat{I}_{HR} \). 
We employ the Adam optimization algorithm~\cite{kingma2014adam}, which is an extension of the stochastic gradient descent \cite{jais2019adam}.

{\bf \emph{A-CNN}}.
To enhance the performance of the network,
we also consider a variant of the $CNN$ architecture where an \emph{attention block} 
\citep{woo2018cbam, fu2019dual} 
is added. The {attention block} enables the network to weigh differently different parts of the input in order to focus on more relevant parts of the data~\citep{zhang2018image,liang2021swinir}. 
The \emph{CNN} network in
Fig.~\ref{fig:CNN_blocks} is  equipped with the attention block and thus it will be referred to as \emph{A-CNN}. 
In Fig.~\ref{fig:CNN_blocks}, after the attention block 
there is one last convolution layer, generating the final
output. 

We will train the \emph{A-CNN} in the same way as the \emph{CNN}. While the goal of adopting \emph{A-CNN} is to improve SR accuracy over a standard $CNN$, we also strive to maintain computational efficiency. In particular, we design the \emph{A-CNN} 
networks so that their number of learnable parameters 
is roughly equivalent. 
In this way, we compare accuracy enabled by the \emph{A-CNN} architecture for a comparable computational cost. We remark that 
the number of learnable parameters
is a standard measure of computational cost. 

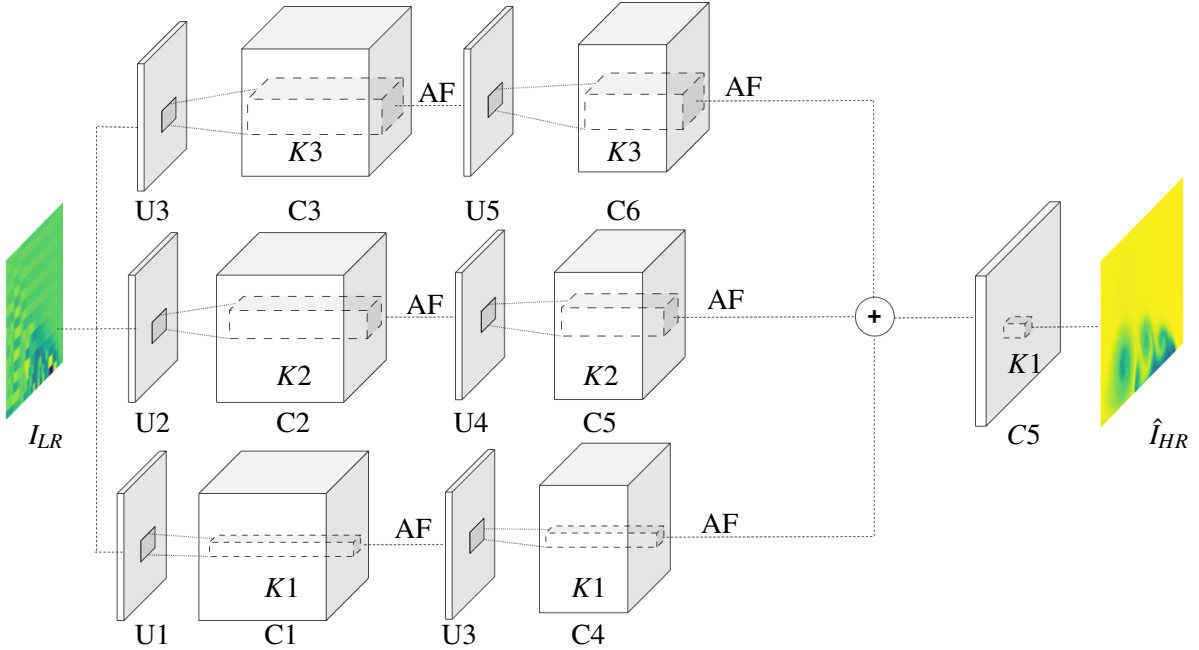
\begin{figure}[htb!]
    \centering
    \begin{overpic}[width=0.95\textwidth]{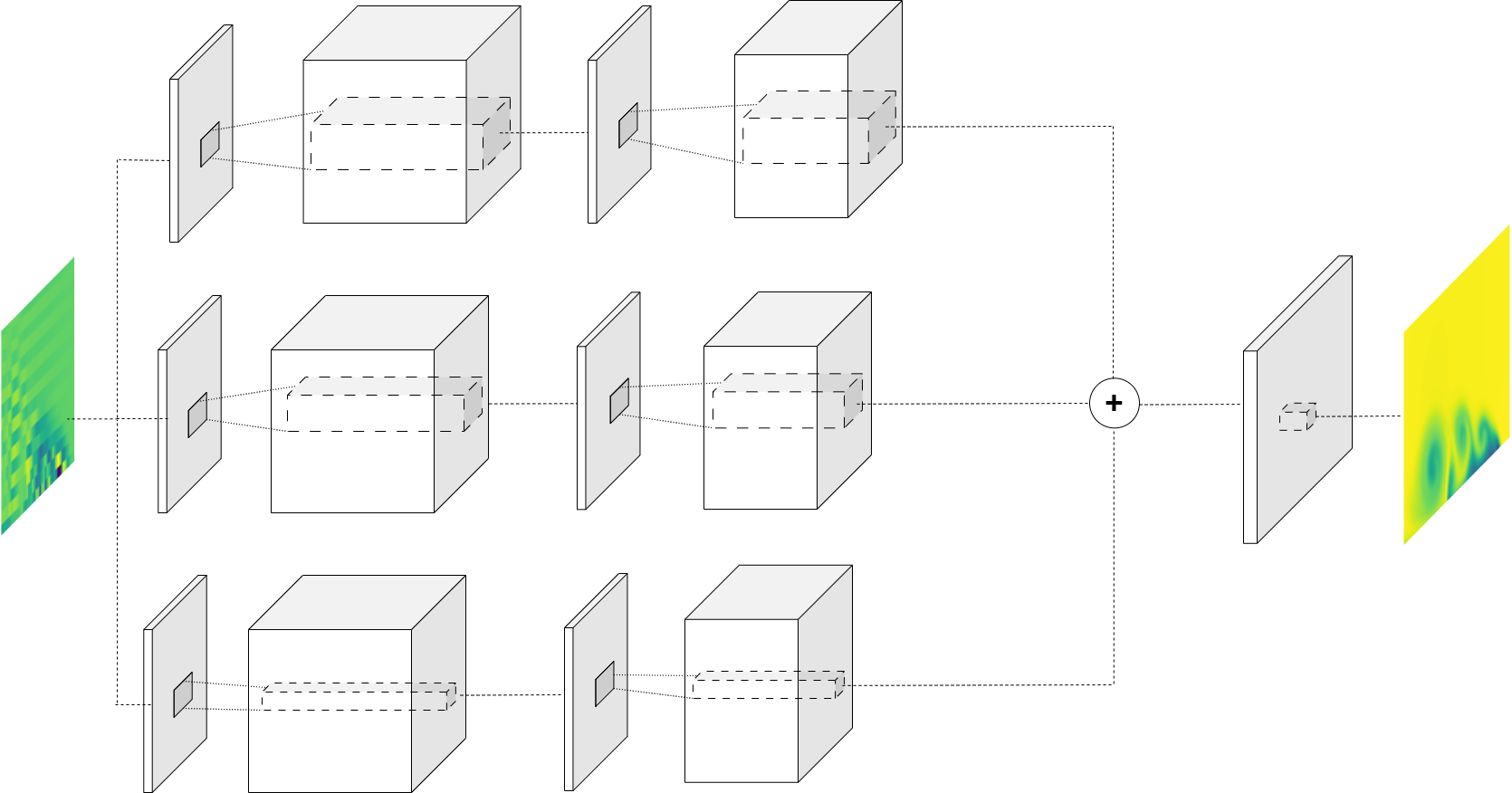}
\put(2,15){$I_{LR}$}
\put(11,-2){U1}
\put(11,16){U2}
\put(11,34){U3}
\put(22,2){{$K1$}}
\put(23,20){{$K2$}}
\put(24,39){{$K3$}}
\put(22,-2){C1}
\put(23,16){C2}
\put(24,34){C3}
\put(33,7){AF}
\put(34,26){AF}
\put(35,44){AF}
\put(37,-2){U3}
\put(38,16){U4}
\put(39,34){U5}
\put(48,-2){C4}
\put(49,16){C5}
\put(51,34){C6}
\put(48,2){$K1$}
\put(49,20){$K2$}
\put(51,39){$K3$}
\put(59,7.5){AF}
\put(59.5,26.5){AF}
\put(61,44.5){AF}
\put(85,15){{$C5$}}
\put(85,21){{$K1$}}
\put(97,15){$\hat{I}_{HR}$}
\end{overpic}
    \caption{Schematic diagram of an \emph{m-CNN} architecture, using the same notation as in Fig.~\ref{fig:CNN_blocks}. 
    }
    \label{fig:CNN_blocks2}
\end{figure}

{\bf \emph{m-CNN}}. We
will also consider a \emph{multi-scale CNN} (\emph{m-CNN}) architecture with three branches
whose convolution layers
employ filters of different size. A schematic of this network is shown in 
Fig.~\ref{fig:CNN_blocks2}. 
This approach is particularly useful when the 
input data are characterized by a wide range of spatial scales 
, as it happens for mesoscale flows.

The \emph{m-CNN} architecture
will be trained in the same way as \emph{CNN} and \emph{A-CNN}. It is also 
designed to have its number of learnable parameters 
is roughly equivalent to 
\emph{CNN} and \emph{A-CNN}.

{\bf \emph{Diff}}.
We consider a diffusion model
because it represents the sate-of-the-art for super-resolution tasks~\citep{xu2025diffusion, huang2025physics22}.
Standard diffusion is a forward process that gradually corrupts a high-resolution image $I_{\text{HR}}$ by adding Gaussian noise $\epsilon$ over a certain number of steps \cite{ho2020denoising}. 
At diffusion step $s\in \{1, \dots, S\}$, 
let $\beta_{s} \in (0,1)$ be the amount of added noise and $\alpha_{s} = 1 - \beta_{s}$
the amount of original information preserved. 
Then, the state 
$X_{s}$ 
of the corruption process can be expressed as:
\begin{equation}
X_{s} = \sqrt{\bar{\alpha}_{s}} I_{\text{HR}} + \sqrt{1 - \bar{\alpha}_{s}} \epsilon, \quad \epsilon \sim \mathcal{N}(0, I)
\label{eq:alpha-noise}
\end{equation}
where $\bar{\alpha}_{s} = \prod_{j=1}^{s} \alpha_s$ is the cumulative product of the noise-scaling factors {and $I$ denotes the identity matrix, ensuring the noise $\epsilon$ is isotropic.} 
We note that a higher index $s$
corresponds to a noisier state of the image.
We recall that, in the case of super-resolution, the input is
a low-resolution image $I_{\text{LR}}$. 
During the training phase of the SR, the $I_{\text{LR}}$ image is 
upsampled to the target (high) resolution and stacked with the noisy $X_s$ along the channel dimension. This combined tensor $[I_{\text{LR}}, X_{s}]$ is passed through a U-Net~\citep{ronneberger2015u,ho2020denoising}, $\epsilon_\theta$, 
trained to predict the noise $\epsilon$ using objective
function:
\begin{equation}
\mathbb{E}_{{s}, I_{\text{HR}}, \epsilon} \left[ \left\| \epsilon - \epsilon_\theta([I_{\text{LR}}, X_{s}], {s}) \right\|_2^2 \right],
\end{equation}
which is the MSE formulated as the expectation $\mathbb{E}$ over
all diffusion steps, training images, and noise samples.
To enhance the robustness of the learned features, we employ an Exponential Moving Average of the model weights, which smooths training fluctuations and improves the generation quality of the final model~\cite{karras2024analyzing}. 
During the testing phase of the SR, the backward (denoising) process begins with pure noise $X_{S} \sim \mathcal{N}(0, I)$. 
To generate the high-resolution output, we remove noise by conditioning each step on $I_{\text{LR}}$ with a 
Conditional Diffusion Model~\cite{saharia2022palette}.
Through a sequence of conditional Gaussian transitions,
this model estimates what the slightly cleaner image 
$X_{s-1}$ should look like, given the current noisy image $X_s$ and the low-resolution guidance $I_{\text{LR}}$.
By stacking the same $I_{\text{LR}}$ conditioning image with the evolving sample $X_s$ at every step $s$, the model ensures the recovered details are consistent with the input~\citep{salimans2022progressive,song2020denoising}. Finally, we utilize a Denoising Diffusion Implicit Model
~\cite{an2023efficient} during inference to accelerate the denoising process 
because it allows for a deterministic mapping and thus 
significantly fewer sampling steps compared to standard stochastic Markovian chains.

\section{Numerical results}\label{Sec:res}
In order to test our super-resolution approach, we consider two standard benchmarks for atmospheric flows: the rising thermal bubble, with the same setting as in \cite{GirfoglioFVCA10}, and the density current, with the same setting as in \cite{girfoglio2025comparative}. 
Both benchmarks involve perturbations of a neutrally stratified atmosphere with uniform background potential temperature. 
In both cases, the potential temperature perturbation computed by the AV model or the Smagorinsky model with a fine mesh is the reference solution. We consider a set of coarser meshes and adopt the techniques described in Sec. \ref{sec:SR} to improve the $\theta'$ field computed with such meshes and the AV model. 

Tab.~\ref{tab:cnn_att_hyperparameters} lists the hyperparameters used for \emph{CNN} and \emph{A-CNN}, while the 
hyperparameters used for \emph{m-CNN}
are reported in Table \ref{tab:cnn_ms_hyperparameters}.
Tab.~\ref{tab:diffusion_cnn_hyperparameters} contains the specifications for \emph{Diff}.
The results for the rising thermal bubble benchmark
are presented in Sec. \ref{sec:RTB}, while Sec. \ref{sec:DC} reports the results for the density current benchmark. We will show in Sec. \ref{sec:RTB} that \emph{CNN} allows to obtain 
accurate results for the rising thermal bubble benchmark, likely due to the relative simplicity of the flow. Hence, for that benchmark we do not consider any of the more sophisticated architectures.
In Sec. \ref{sec:DC}, we will show that \emph{CNN}
fails to provide accurate results as the flow
is more complex and thus we will focus on 
\emph{A-CNN}, \emph{m-CNN}, and \emph{Diff}.

For the numerical solution of both benchmarks, we used
GEA\footnote{The source code of GEA can be found in \url{https://github.com/GEA-Geophysical-and-Environmental-Apps/GEA}} - Geophysical and Environmental Applications \cite{GEA,GIRFOGLIO2025106510, GQR_OF_clima}, an open-source software package for atmospheric and oceanic modeling based on
the OpenFOAM library \cite{openfoam}.



\begin{table}[htb!]
\centering
\begin{tabular}{|c|c|}
\hline
\textbf{No. of Layers}         & 3                            \\ \hline
\textbf{No. of Filters}        & \{32, 64, 1\}              \\ \hline
\textbf{Kernel Size}              & 3x3                          \\ \hline
\textbf{Batch Size} & 2 \\ \hline
\textbf{Upsampling method}        & Bilinear Interpolation                   \\ \hline
\textbf{Epochs}                   & 2000                         \\ \hline
\textbf{Activation Function}      & ReLU                         \\ \hline
\textbf{Loss Function}            & MSE                          \\ \hline
\textbf{Optimizer}                & Adam                         \\ \hline
\textbf{Learning Rate}            & 1E-04                        \\ \hline
\textbf{No. of  Attention Blocks} ($A$-$CNN$ only)         & 2                            \\ \hline
\end{tabular}
\caption{Hyperparameters of the $CNN$ and $A$-$CNN$ networks. 
}

\label{tab:cnn_att_hyperparameters}
\end{table}


\begin{table}[htb!]
\centering
\begin{tabular}{|c|c|}
\hline
\textbf{No. of Scales} & 3 \\ \hline
\textbf{No. of Layers} & 7 \\ \hline
\textbf{No. of Filters per Layer} & \{32, 64, 1\} \\ \hline
\textbf{Kernel Sizes} & \{3x3, 5x5, 7x7\} \\ \hline
\textbf{Batch Size} & 2 \\ \hline
\textbf{Upsampling Method} & Bilinear Interpolation \\ \hline
\textbf{Epochs}                   & 2000 \\ \hline
\textbf{Activation Function} & ReLU \\ \hline
\textbf{Loss Function} & MSE \\ \hline
\textbf{Optimizer} & Adam \\ \hline
\textbf{Learning Rate} & 1E-04 \\ \hline
\end{tabular}
\caption{Hyperparameters of the $m$-$CNN$ network. 
}
\label{tab:cnn_ms_hyperparameters}
\end{table}

\begin{table}[htb!]
\centering
\begin{tabular}{|c|c|}
\hline
\textbf{U-Net Encoder Blocks} & 3 \\ \hline
\textbf{Filters per Layer} & {[64, 128, 256]} \\ \hline
\textbf{Kernel Size} & 3x3 (Conv2d) \\ \hline
\textbf{Diffusion Noise Steps} & 1000 (DDPM Training) \\ \hline
\textbf{Inference Sampling} & DDIM (50 steps) \\ \hline
\textbf{Batch Size} & 2 \\ \hline
\textbf{Upsampling Method} & Bicubic Interpolation \\ \hline
\textbf{Epochs} & 200 \\ \hline
\textbf{Activation Function} & GELU \\ \hline
\textbf{Loss Function} & MSE (Noise Prediction) \\ \hline
\textbf{Optimizer} & Adam \\ \hline
\textbf{Learning Rate} & 1E-04 \\ \hline
\textbf{Moving Average} & EMA (Exponential Moving Average) \\ \hline
\end{tabular}
\caption{Hyperparameters of the \emph{Diff} network.}
\label{tab:diffusion_cnn_hyperparameters}
\end{table}

\subsection{Rising thermal bubble}\label{sec:RTB}
The computational domain in the $xz$-plane is $\Omega=[0, 5000]\times[0, 10000]$ m$^2$ and the time interval of interest is $(0, 1020]$ s. 
The initial density is given by \begin{align}
\rho^0 = \frac{p_g}{R \theta_0} \left(\frac{p}{p_g}\right)^{c_{v}/c_p}, \quad \text{with} \ \ p = p_g \left( 1 - \frac{g z}{c_p \theta^0} \right)^{c_p/R}, \label{eq:rho_wb}
\end{align}
with 
$c_p = R + c_v$, $c_v = 715.5$ J/(Kg K) and $R = 287$ J/(Kg K).
In \eqref{eq:rho_wb}, $\theta^0$ is the initial potential temperature, which is defined as:
\begin{equation}
\theta^0 = 300 + 2\left[ 1 - \frac{r}{r_0} \right] ~ \textrm{if $r\leq r_0=2000~\mathrm{m}$}, \quad\theta^0 = 300
~ \textrm{otherwise},
\label{warmEqn1}
\end{equation}
where $r = \sqrt[]{(x-x_{c})^{2} + (z-z_{c})^{2}}$ and $(x_c,z_c) = (5000,2000)~\mathrm{m}$ \cite{ahmadLindeman2007,ahmad2018}.
Eq.~\eqref{eq:rho_wb}-\eqref{warmEqn1} represent a neutrally stratified atmosphere with uniform background potential temperature at $ 300~\mathrm{K}$ perturbed by a circular bubble of warmer air. 
The initial velocity field is zero everywhere and the initial specific enthalpy is given by:
\begin{align}
h^{0} = c_p \theta^0 \left( \frac{p}{p_g} \right)^{\frac{R}{c_{p}}}.
\label{eq:e0}
\end{align}
We impose impenetrable, free-slip boundary conditions on all the boundary. 



As mentioned at the beginning pf Sec.~\ref{Sec:res}, we will consider only the super-resolution with the \emph{CNN} architecture for this benchmark. First, the reference solution will be the potential temperature perturbation computed by the AV model, sampled $40,000$ times (i.e., the solution computed at every time step).
Then, the reference solution will become the potential temperature perturbation computed by Smagorinsky model, sampled 
$10,200$ times (i.e., one computed solution every 8 time steps). 
This difference in the sampling is due to a
trade-off between accuracy and computational cost. In fact, the high-resolution simulation
with the Smagorinsky model require a finer
mesh and a smaller time step than the 
high-resolution simulation
with the AV model in order to resolve finer spatial and temporal scales that ge smoother out by the AV model. Thus, each sample from the Smagorinsky model is a larger amount of data and using $40,000$ samples for it was beyond the computational facilities available to the authors. 
From the results presented later, it seems that $10,200$ samples from the Smagorinsky model is enough to accurately capture the flow evolution while keeping the computational cost reasonable. 
Following a conventional SR approach, the solution snapshots are randomly shuffled and split for training and testing. Let us initially 
take 80\% of the snapshots for training and leave
the remaining 20\% for validation. Since 80\% is a high percentage, later we will reduce it. 
We recall that all hyperparameters for the \emph{CNN} network are found in Tab.~\ref{tab:cnn_att_hyperparameters}.

\begin{figure}[htb!]
    \centering
\begin{overpic}[width=0.49\textwidth,grid=false]{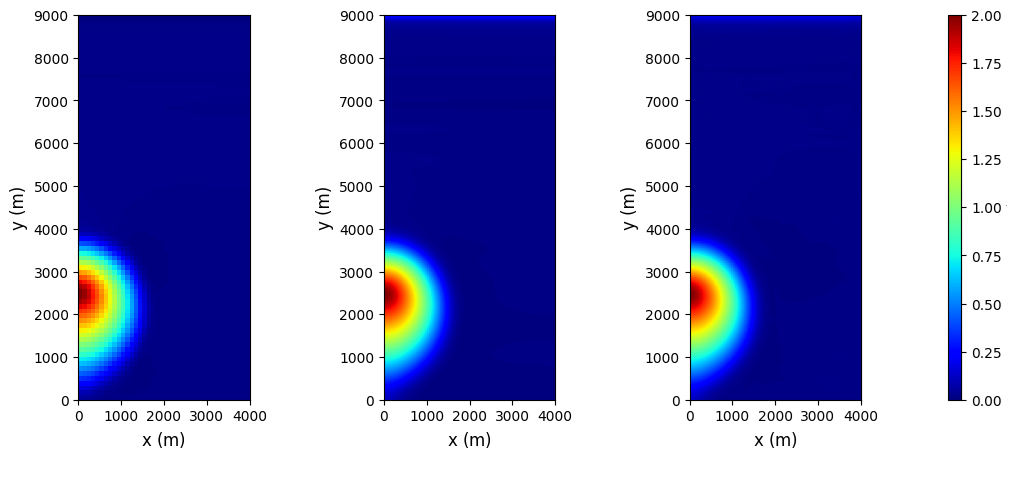}
        \put(8,40){\scriptsize{\textcolor{white}{$t/T=0.25$}}}
        \put(39,40){\scriptsize{\textcolor{white}{$t/T=0.25$}}}
        \put(69,40){\scriptsize{\textcolor{white}{$t/T=0.25$}}}
        \put(13,47){\footnotesize{$I_{LR}$}}
        \put(41,47){\footnotesize{$I_{HR}$}}
        \put(63,47){\footnotesize{SR with \emph{CNN}}}
        \put(91,47){\footnotesize{$\theta'$}}
    \end{overpic} 
\begin{overpic}[width=0.49\textwidth,grid=false]{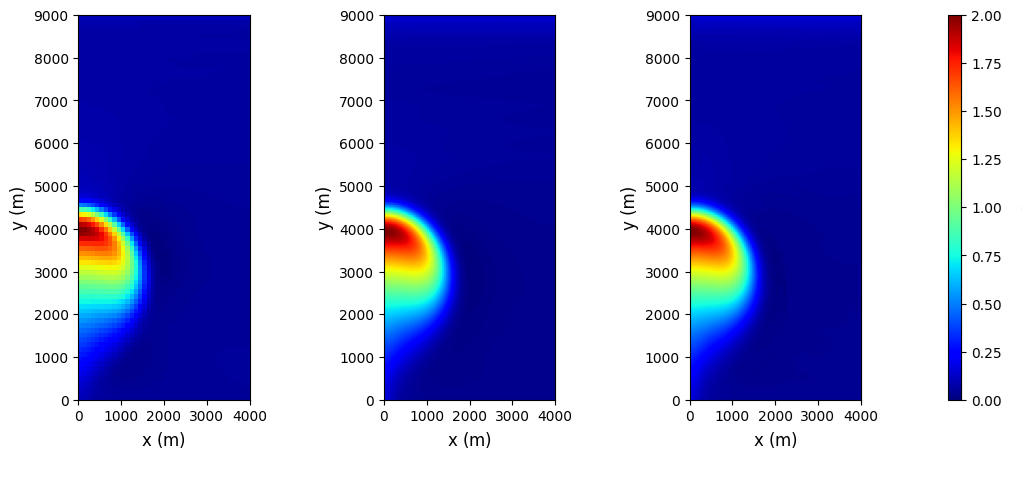}
        \put(9,40){\scriptsize{\textcolor{white}{$t/T=0.5$}}}
        \put(39,40){\scriptsize{\textcolor{white}{$t/T=0.5$}}}
        \put(70,40){\scriptsize{\textcolor{white}{$t/T=0.5$}}}
                \put(13,47){\footnotesize{$I_{LR}$}}
        \put(41,47){\footnotesize{$I_{HR}$}}
        \put(63,47){\footnotesize{SR with \emph{CNN}}}
        \put(91,47){\footnotesize{$\theta'$}}
    \end{overpic}\\
    \vskip .2cm
\begin{overpic}[width=0.49\textwidth,grid=false]{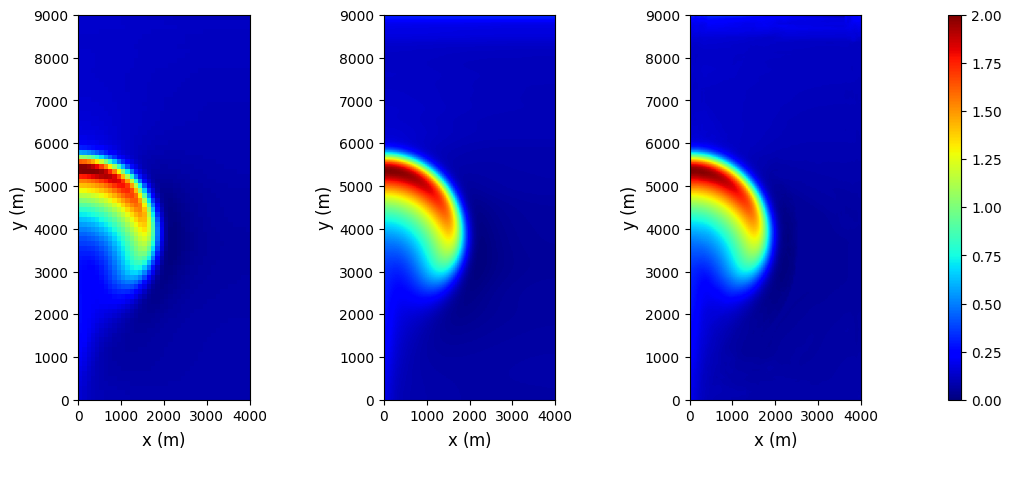}
                \put(8,40){\scriptsize{\textcolor{white}{$t/T=0.75$}}}
                \put(38,40){\scriptsize{\textcolor{white}{$t/T=0.75$}}}
        \put(67.3,40){\scriptsize{\textcolor{white}{$t/T=0.75$}}}
        \put(13,47){\footnotesize{$I_{LR}$}}
        \put(43,47){\footnotesize{$I_{HR}$}}
        \put(64,47){\footnotesize{SR with \emph{CNN}}}
        \put(91,47){\footnotesize{$\theta'$}}
    \end{overpic} 
 \begin{overpic}[width=0.49\textwidth,grid=false]{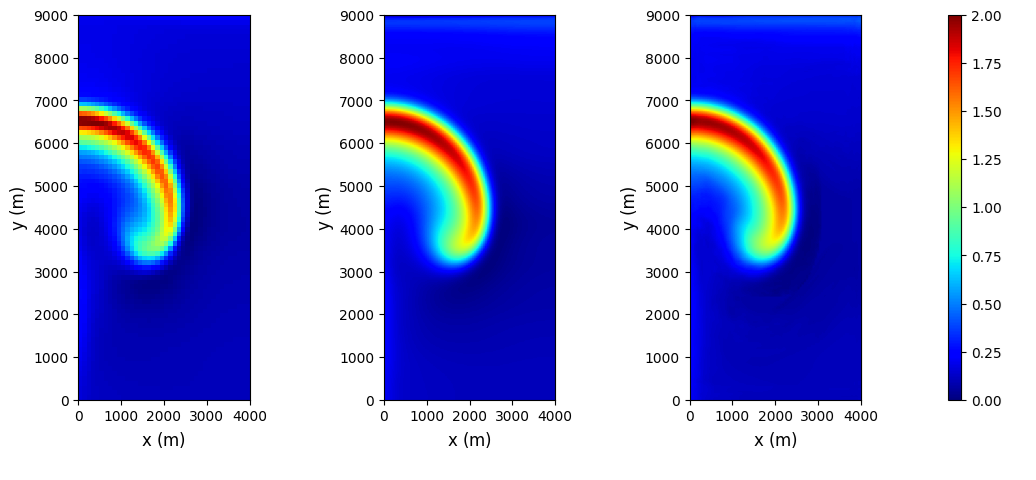}
        \put(10,40){\scriptsize{\textcolor{white}{$t/T=1$}}}
        \put(40,40){\scriptsize{\textcolor{white}{$t/T=1$}}}
         \put(70,40){\scriptsize{\textcolor{white}{$t/T=1$}}}
        \put(13,47){\footnotesize{$I_{LR}$}}
        \put(43,47){\footnotesize{$I_{HR}$}}
        \put(65,47){\footnotesize{SR with \emph{CNN}}}
        \put(91,47){\footnotesize{$\theta'$}}
    \end{overpic}
    \caption{Rising bubble, coarse mesh 
    $h = 125$ m: low-resolution solution computed by the AV model (left in each panel), reference solution computed by the AV model (center in each panel), and improvement by the super resolution with \emph{CNN} (right in each panel) for different times $t$, with $T = 1020$ s.
    }
    \label{fig:125_62}
\end{figure}

\begin{figure}[htb!]
    \centering
\begin{overpic}[width=0.49\textwidth,grid=false]{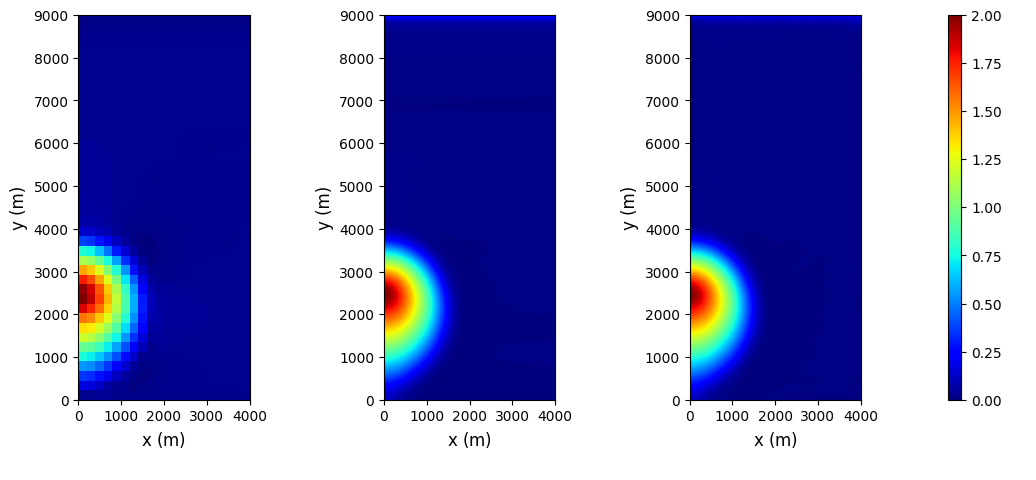}
        \put(8,40){\scriptsize{\textcolor{white}{$t/T=0.25$}}}
        \put(38,40){\scriptsize{\textcolor{white}{$t/T=0.25$}}}
        \put(67,40){\scriptsize{\textcolor{white}{$t/T=0.25$}}}
        \put(13,47){\footnotesize{$I_{LR}$}}
        \put(41,47){\footnotesize{$I_{HR}$}}
        \put(63,47){\footnotesize{SR with \emph{CNN}}}
        \put(91,47){\footnotesize{$\theta'$}}
    \end{overpic} 
\begin{overpic}[width=0.49\textwidth,grid=false]{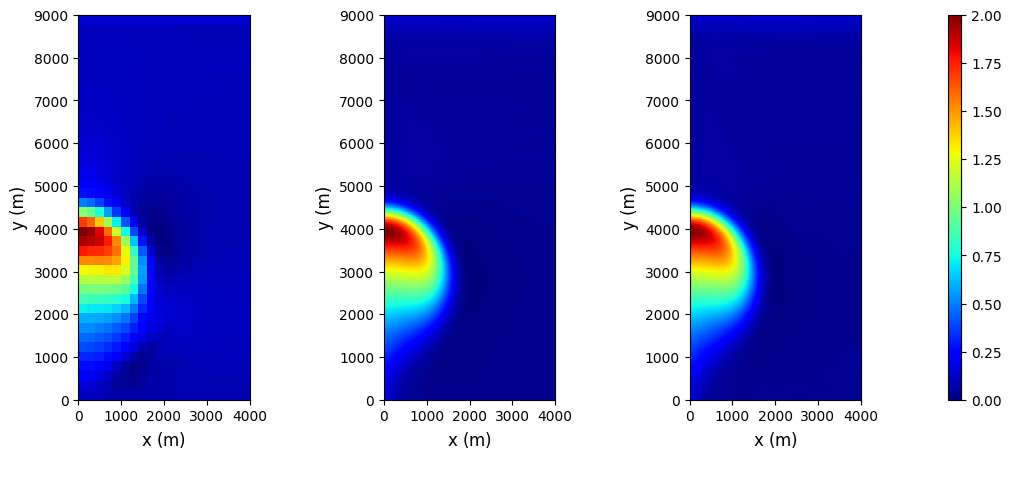}
         \put(9,40){\scriptsize{\textcolor{white}{$t/T=0.5$}}}
        \put(38,40){\scriptsize{\textcolor{white}{$t/T=0.5$}}}
        \put(68,40){\scriptsize{\textcolor{white}{$t/T=0.5$}}}
                \put(13,47){\footnotesize{$I_{LR}$}}
        \put(41,47){\footnotesize{$I_{HR}$}}
        \put(63,47){\footnotesize{SR with \emph{CNN}}}
        \put(91,47){\footnotesize{$\theta'$}}
    \end{overpic} \\
    \vskip .2cm
\begin{overpic}[width=0.49\textwidth,grid=false]{Figures/250_62_t50_tolto.png}
        \put(8,40){\scriptsize{\textcolor{white}{$t/T=0.75$}}}
        \put(37.5,40){\scriptsize{\textcolor{white}{$t/T=0.75$}}}
        \put(67.5,40){\scriptsize{\textcolor{white}{$t/T=0.75$}}}
                \put(13,47){\footnotesize{$I_{LR}$}}
        \put(41,47){\footnotesize{$I_{HR}$}}
        \put(63,47){\footnotesize{SR with \emph{CNN}}}
        \put(91,47){\footnotesize{$\theta'$}}
    \end{overpic}
\begin{overpic}[width=0.49\textwidth,grid=false]{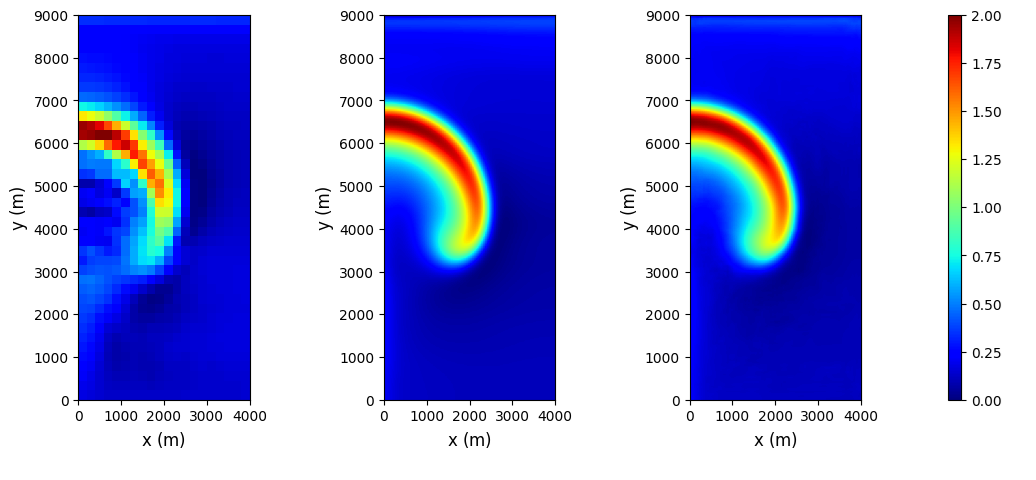}
        \put(10,40){\scriptsize{\textcolor{white}{$t/T=1$}}}
        \put(40,40){\scriptsize{\textcolor{white}{$t/T=1$}}}
         \put(70,40){\scriptsize{\textcolor{white}{$t/T=1$}}}
        \put(13,47){\footnotesize{$I_{LR}$}}
        \put(41,47){\footnotesize{$I_{HR}$}}
        \put(63,47){\footnotesize{SR with \emph{CNN}}}
        \put(91,47){\footnotesize{$\theta'$}}
    \end{overpic}
    \caption{Rising bubble, coarse mesh 
    $h = 250$ m: low-resolution solution computed by the AV model (left in each panel), reference solution computed by the AV model (center in each panel), and improvement by the super resolution with \emph{CNN} (right in each panel) for different times $t$, with $T = 1020$ s.}
    \label{fig:250_62}
\end{figure}

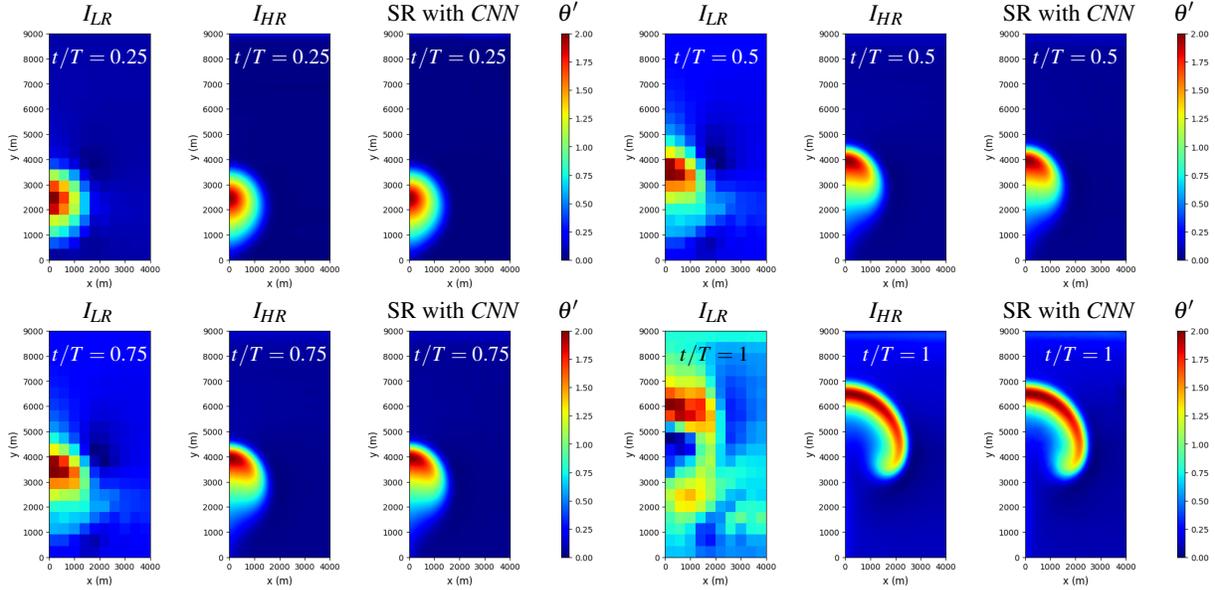
\begin{figure}[htb!]
     \centering
\begin{overpic}[width=0.49\textwidth,grid=false]{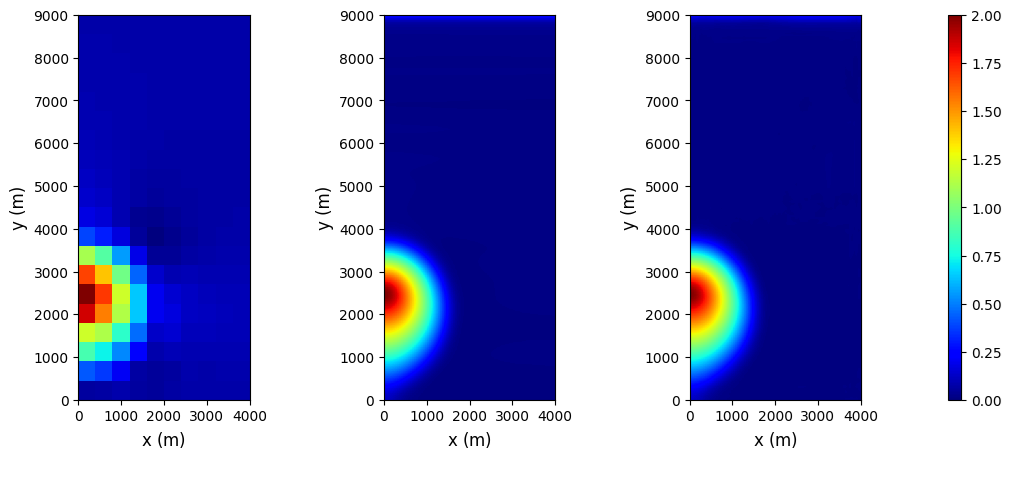}
        \put(8,40){\scriptsize{\textcolor{white}{$t/T=0.25$}}}
        \put(38,40){\scriptsize{\textcolor{white}{$t/T=0.25$}}}
        \put(67,40){\scriptsize{\textcolor{white}{$t/T=0.25$}}}
        \put(13,47){\footnotesize{$I_{LR}$}}
        \put(41,47){\footnotesize{$I_{HR}$}}
        \put(63,47){\footnotesize{SR with \emph{CNN}}}
        \put(91,47){\footnotesize{$\theta'$}}
    \end{overpic} 
\begin{overpic}[width=0.49\textwidth,grid=false]{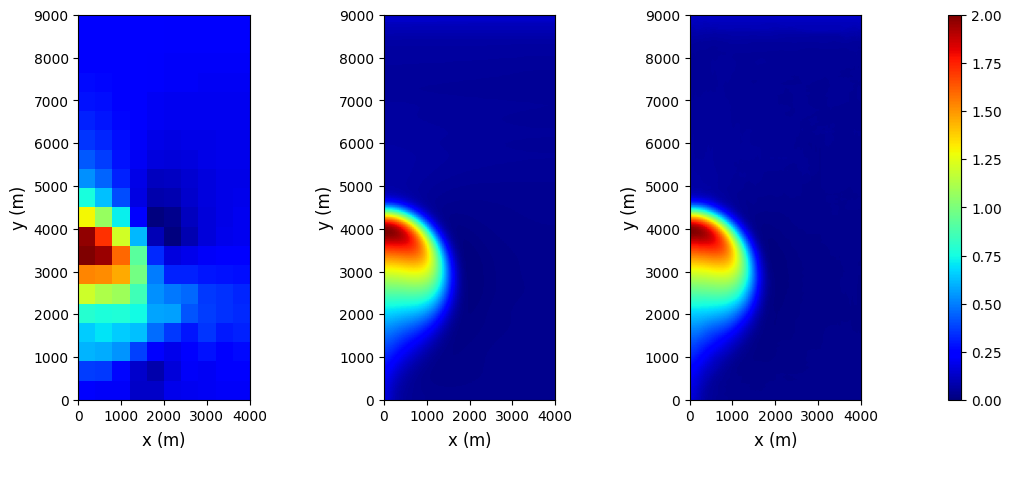}
         \put(9,40){\scriptsize{\textcolor{white}{$t/T=0.5$}}}
        \put(38,40){\scriptsize{\textcolor{white}{$t/T=0.5$}}}
        \put(68,40){\scriptsize{\textcolor{white}{$t/T=0.5$}}}
                \put(13,47){\footnotesize{$I_{LR}$}}
        \put(41,47){\footnotesize{$I_{HR}$}}
        \put(63,47){\footnotesize{SR with \emph{CNN}}}
        \put(91,47){\footnotesize{$\theta'$}}
    \end{overpic}\\
    \vskip .2cm
\begin{overpic}[width=0.49\textwidth,grid=false]{Figures/500_62_5_t50_tolto.png}
        \put(8,40){\scriptsize{\textcolor{white}{$t/T=0.75$}}}
        \put(37.5,40){\scriptsize{\textcolor{white}{$t/T=0.75$}}}
        \put(67.5,40){\scriptsize{\textcolor{white}{$t/T=0.75$}}}
                \put(13,47){\footnotesize{$I_{LR}$}}
        \put(41,47){\footnotesize{$I_{HR}$}}
        \put(63,47){\footnotesize{SR with \emph{CNN}}}
        \put(91,47){\footnotesize{$\theta'$}}
    \end{overpic} 
\begin{overpic}[width=0.49\textwidth,grid=false]{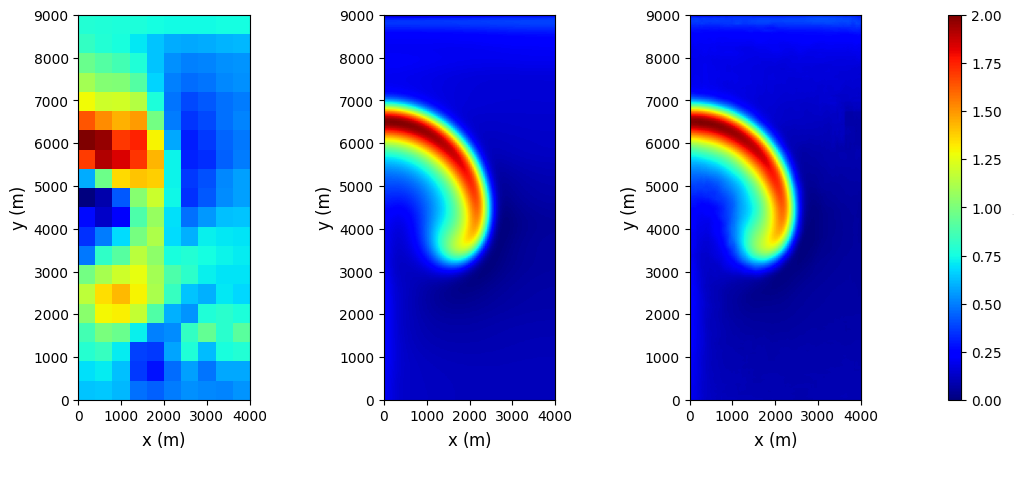}
        \put(10,40){\scriptsize{{$t/T=1$}}}
        \put(40,40){\scriptsize{\textcolor{white}{$t/T=1$}}}
         \put(70,40){\scriptsize{\textcolor{white}{$t/T=1$}}}
        \put(13,47){\footnotesize{$I_{LR}$}}
        \put(41,47){\footnotesize{$I_{HR}$}}
        \put(63,47){\footnotesize{SR with \emph{CNN}}}
        \put(91,47){\footnotesize{$\theta'$}}
    \end{overpic}
    
    \caption{
    Rising bubble, coarse mesh 
    $h = 500$ m: low-resolution solution computed by the AV model (left in each panel), reference solution computed by the AV model (center in each panel), and improvement by the super resolution with \emph{CNN} (right in each panel) for different times $t$, with $T = 1020$ s. 
    }
    \label{fig:500_62}
\end{figure}

Let us start by considering as high-resolution dataset  $I_{HR}$ the solutions provided by the AV model with a structured mesh with grid size $h = 62.5$ m and time $\Delta t = 0.0255$ s.
The low-resolution data $I_{LR}$ are the solutions computed by the same model with three coarser structured meshes with mesh size $h=125, 250, 500$ m, with the same time step ($\Delta t = 0.0255$ s).
Fig. \ref{fig:125_62} and \ref{fig:250_62} compare the reference solutions in $I_{HR}$ with the solutions 
obtained with coarse meshes $h = 125$ m and $h = 250$ m, respectively, and their improvement through the standard $CNN$ architecture. 
Note that the low-resolution solutions computed with both meshes $h = 125$ m and $h = 250$ m (left in each panel in Fig. \ref{fig:125_62} and \ref{fig:250_62}) feature
unphysical oscillations, which are completely cured by
the super-resolution with $CNN$ (right in each panel in Fig. \ref{fig:125_62} and \ref{fig:250_62}). 
The end result
in both cases is a solution that compares very well with
the reference solution (center in each panel in Fig. \ref{fig:125_62} and \ref{fig:250_62}). 
When mesh $h = 500$ m is adopted, the low-resolution solution is even worse in terms of accuracy. See Fig.~\ref{fig:500_62}
(left in each panel). 
Nonetheless, the super resolution with $CNN$ is able to produce an accurate solution. Compare the center and right images in each panel in Fig.~\ref{fig:500_62}.

\begin{figure}[htb!]
    \centering
\begin{overpic}[width=0.55\textwidth,grid=false]{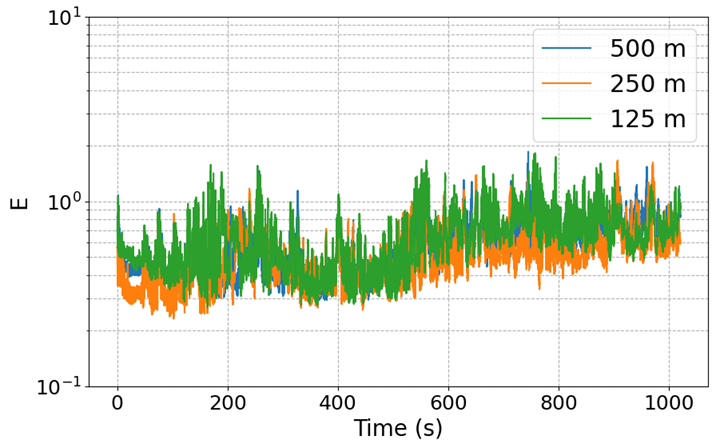}
    \end{overpic}    
    \caption{Rising bubble, AV model for the reference solution: evolution of error \eqref{eq:L2_norm} for the solutions computed by the AV model with coarse meshes  $h = 125, 250, 500$ m. 
    }
    \label{fig:L2_norm_AV}
\end{figure}

In order to provide a more quantitative comparison related to Fig.~\ref{fig:125_62}-\ref{fig:500_62}, Fig.~\ref{fig:L2_norm_AV} shows the time evolution of the $L^2$-norm relative error $E$ defined as: 
\begin{equation}
E = \dfrac{||{SR}(\theta'_{LR}) - \theta'_{HR}||_{L^2}}{||\theta'_{HR}||_{L^2}}\times 100
\label{eq:L2_norm}
\end{equation}
We observe that the relative error does not exceed 2\%, even in the case of the coarsest mesh $h = 500$ m. 

\begin{figure}[htb!]
    \centering
\begin{overpic}[width=0.49\textwidth,grid=false]{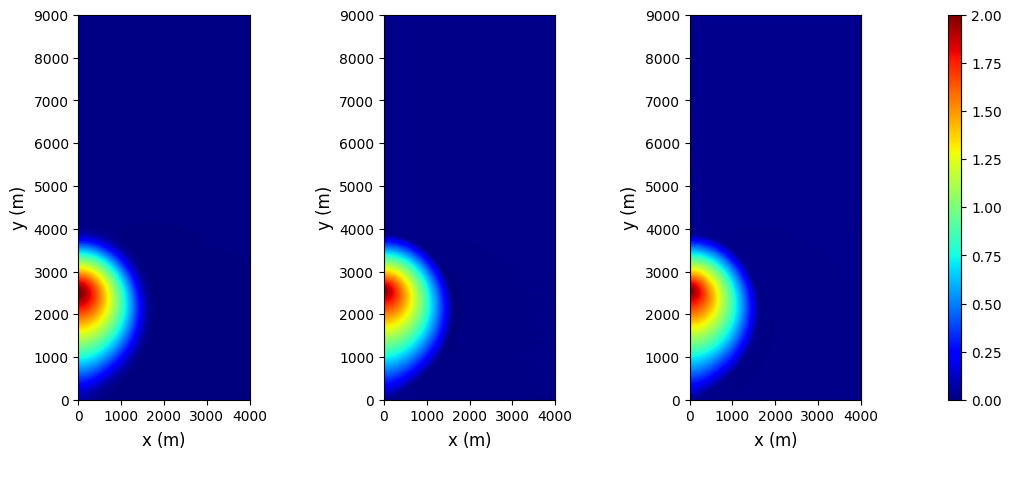}
        \put(8,40){\scriptsize{\textcolor{white}{$t/T=0.25$}}}
        \put(38,40){\scriptsize{\textcolor{white}{$t/T=0.25$}}}
        \put(67,40){\scriptsize{\textcolor{white}{$t/T=0.25$}}}
        \put(13,47){\footnotesize{$I_{LR}$}}
        \put(42,47){\footnotesize{$I_{HR}$}}
        \put(63,47){\footnotesize{SR with \emph{CNN}}}
        \put(91,47){\footnotesize{$\theta'$}}
    \end{overpic} 
\begin{overpic}[width=0.49\textwidth,grid=false]{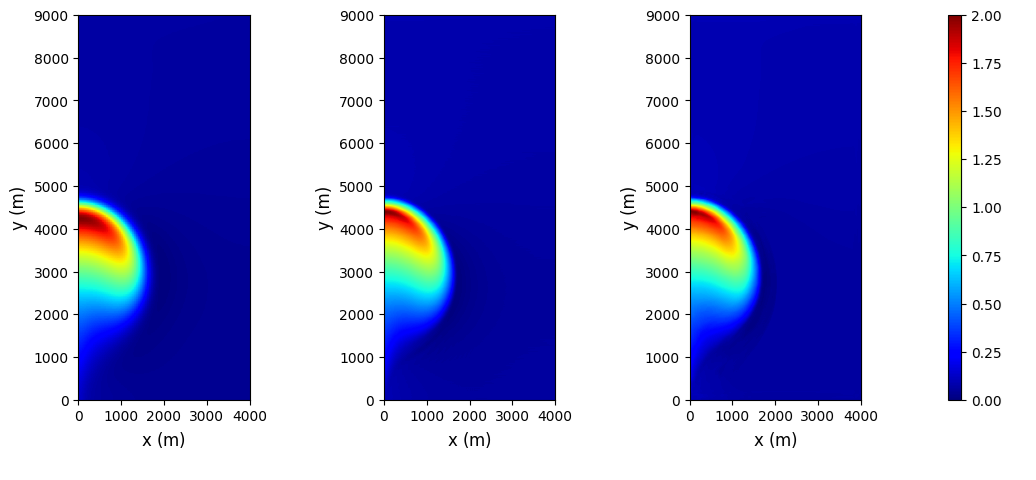}
         \put(9,40){\scriptsize{\textcolor{white}{$t/T=0.5$}}}
        \put(38,40){\scriptsize{\textcolor{white}{$t/T=0.5$}}}
        \put(68,40){\scriptsize{\textcolor{white}{$t/T=0.5$}}}
        \put(13,47){\footnotesize{$I_{LR}$}}
        \put(42,47){\footnotesize{$I_{HR}$}}
        \put(63,47){\footnotesize{SR with \emph{CNN}}}
        \put(91,47){\footnotesize{$\theta'$}}
    \end{overpic}\\
    \vskip .2cm
\begin{overpic}[width=0.49\textwidth,grid=false]{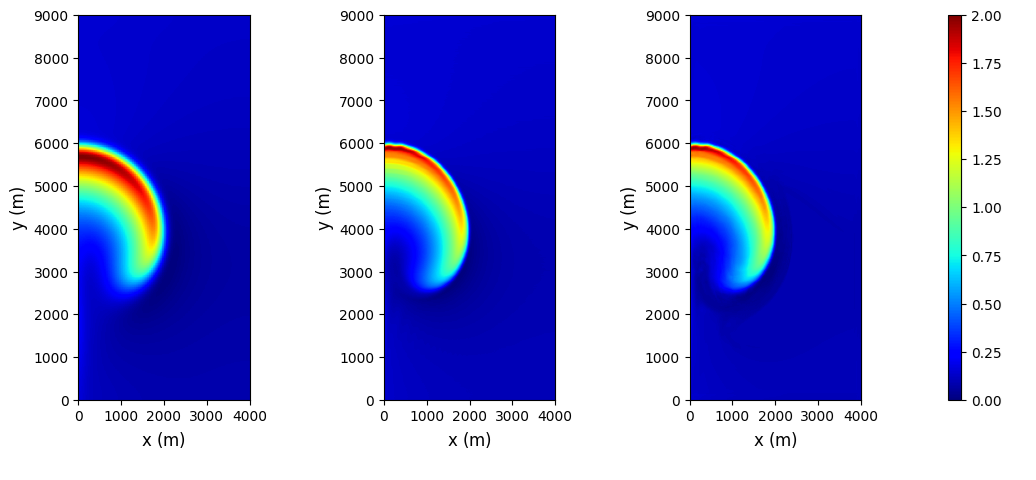}
        \put(8,40){\scriptsize{\textcolor{white}{$t/T=0.75$}}}
        \put(37.5,40){\scriptsize{\textcolor{white}{$t/T=0.75$}}}
        \put(67.5,40){\scriptsize{\textcolor{white}{$t/T=0.75$}}}
        \put(13,47){\footnotesize{$I_{LR}$}}
        \put(42,47){\footnotesize{$I_{HR}$}}
        \put(63,47){\footnotesize{SR with \emph{CNN}}}
        \put(91,47){\footnotesize{$\theta'$}}
    \end{overpic} 
\begin{overpic}[width=0.49\textwidth,grid=false]{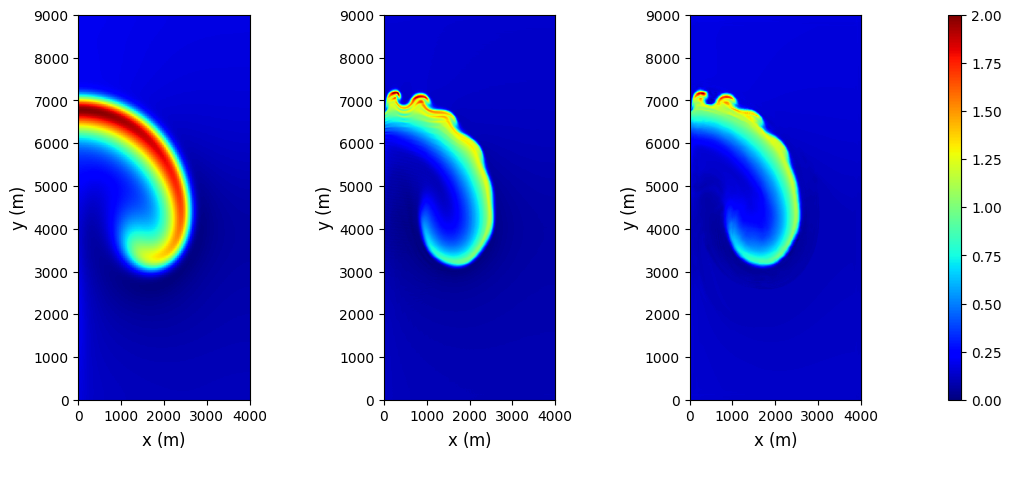}
        \put(10,40){\scriptsize{\textcolor{white}{$t/T=1$}}}
        \put(40,40){\scriptsize{\textcolor{white}{$t/T=1$}}}
         \put(70,40){\scriptsize{\textcolor{white}{$t/T=1$}}}
        \put(13,47){\footnotesize{$I_{LR}$}}
        \put(42,47){\footnotesize{$I_{HR}$}}
        \put(63,47){\footnotesize{SR with \emph{CNN}}}
        \put(91,47){\footnotesize{$\theta'$}}
    \end{overpic}
    
    \caption{Rising bubble, coarse mesh 
    $h = 62$ m: low-resolution solution computed by the AV model (left in each panel), reference solution by the Smagorinsky model (center in each panel), and
    improvement by the super resolution with \emph{CNN} (right in each panel)  for different times $t$, with $T = 1020$ s. 
    }
    \label{fig:62_32_LES}
\end{figure}

\begin{figure}[htb!]
    \centering
\begin{overpic}[width=0.49\textwidth,grid=false]{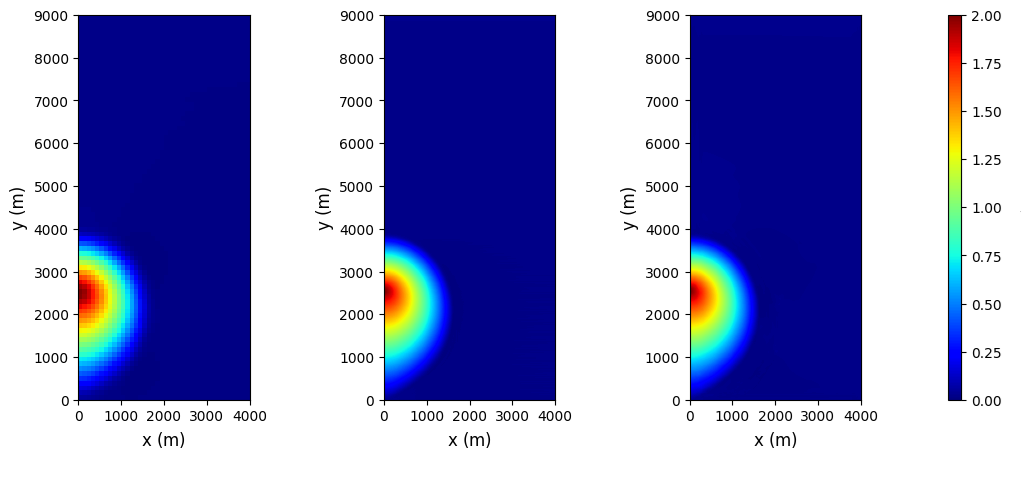}
        \put(8,40){\scriptsize{\textcolor{white}{$t/T=0.25$}}}
        \put(38,40){\scriptsize{\textcolor{white}{$t/T=0.25$}}}
        \put(67,40){\scriptsize{\textcolor{white}{$t/T=0.25$}}}
        \put(13,47){\footnotesize{$I_{LR}$}}
        \put(41,47){\footnotesize{$I_{HR}$}}
        \put(63,47){\footnotesize{SR with \emph{CNN}}}
        \put(91,47){\footnotesize{$\theta'$}}
    \end{overpic} 
\begin{overpic}[width=0.49\textwidth,grid=false]{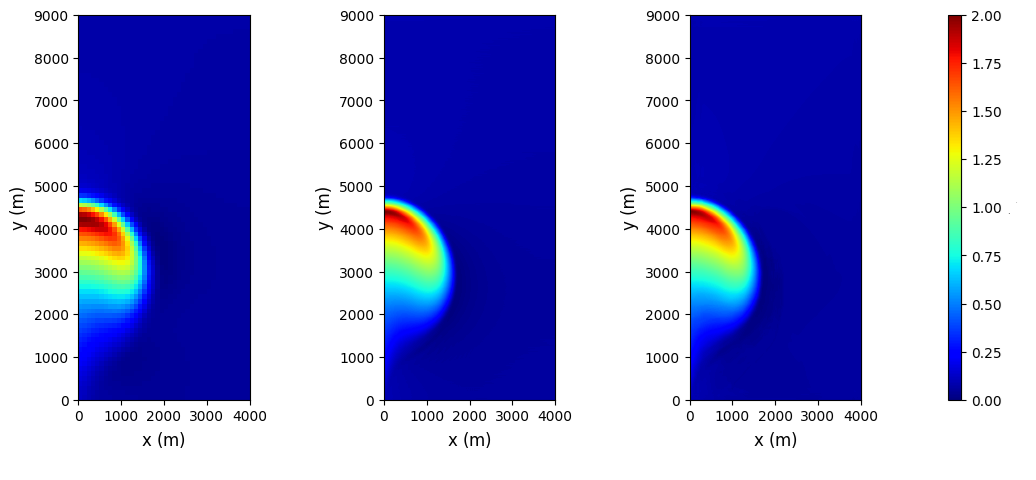}
         \put(9,40){\scriptsize{\textcolor{white}{$t/T=0.5$}}}
        \put(38,40){\scriptsize{\textcolor{white}{$t/T=0.5$}}}
        \put(68,40){\scriptsize{\textcolor{white}{$t/T=0.5$}}}
                \put(13,47){\footnotesize{$I_{LR}$}}
        \put(41,47){\footnotesize{$I_{HR}$}}
        \put(63,47){\footnotesize{SR with \emph{CNN}}}
        \put(91,47){\footnotesize{$\theta'$}}
    \end{overpic}\\
    \vskip .2cm
\begin{overpic}[width=0.49\textwidth,grid=false]{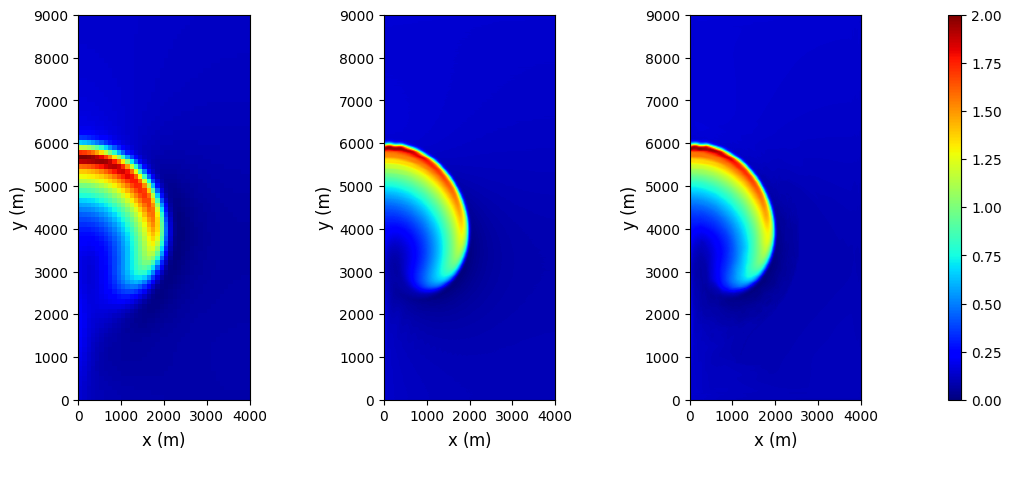}
        \put(8,40){\scriptsize{\textcolor{white}{$t/T=0.75$}}}
        \put(37.5,40){\scriptsize{\textcolor{white}{$t/T=0.75$}}}
        \put(67.5,40){\scriptsize{\textcolor{white}{$t/T=0.75$}}}
                \put(13,47){\footnotesize{$I_{LR}$}}
        \put(41,47){\footnotesize{$I_{HR}$}}
        \put(63,47){\footnotesize{SR with \emph{CNN}}}
        \put(91,47){\footnotesize{$\theta'$}}
    \end{overpic}
\begin{overpic}[width=0.49\textwidth,grid=false]{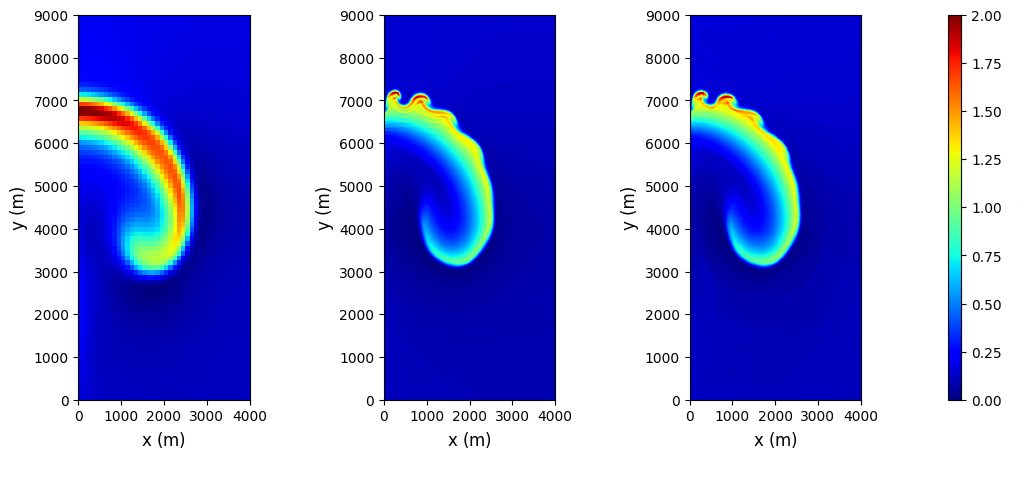}
        \put(10,40){\scriptsize{\textcolor{white}{$t/T=1$}}}
        \put(40,40){\scriptsize{\textcolor{white}{$t/T=1$}}}
         \put(70,40){\scriptsize{\textcolor{white}{$t/T=1$}}}
        \put(13,47){\footnotesize{$I_{LR}$}}
        \put(41,47){\footnotesize{$I_{HR}$}}
        \put(63,47){\footnotesize{SR with \emph{CNN}}}
        \put(91,47){\footnotesize{$\theta'$}}
    \end{overpic}
    
    \caption{Rising bubble, coarse mesh 
    $h = 125$ m: low-resolution solution computed by the AV model (left in each panel), reference solution computed by the Smagorinsky model (center in each panel), and improvement by the super resolution with \emph{CNN} (right in each panel) for different times $t$, with $T = 1020$ s. }
    \label{fig:125_32_LES}
\end{figure}

\begin{figure}[htb!]
   \centering
\begin{overpic}[width=0.49\textwidth,grid=false]{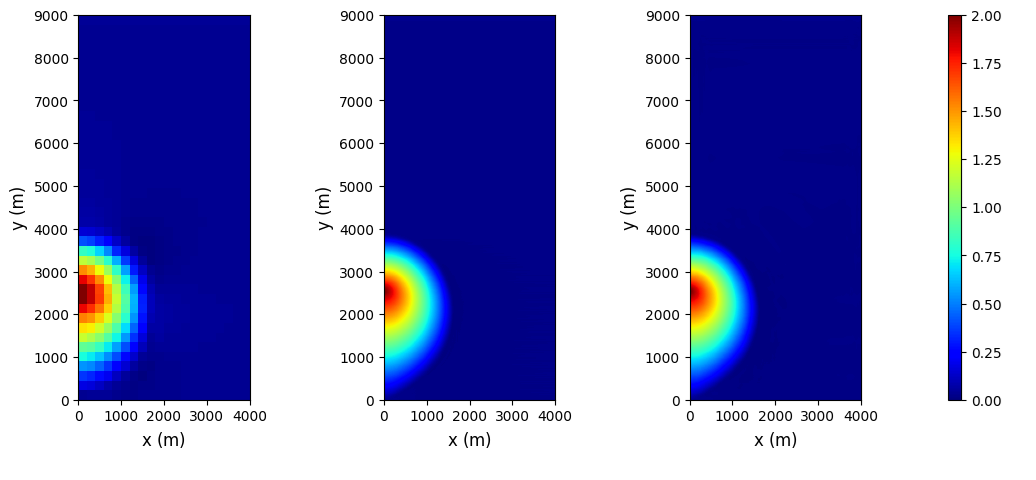}
        \put(8,40){\scriptsize{\textcolor{white}{$t/T=0.25$}}}
        \put(38,40){\scriptsize{\textcolor{white}{$t/T=0.25$}}}
        \put(67,40){\scriptsize{\textcolor{white}{$t/T=0.25$}}}
        \put(13,47){\footnotesize{$I_{LR}$}}
        \put(41,47){\footnotesize{$I_{HR}$}}
        \put(63,47){\footnotesize{SR with \emph{CNN}}}
        \put(91,47){\footnotesize{$\theta'$}}
    \end{overpic} 
\begin{overpic}[width=0.49\textwidth,grid=false]{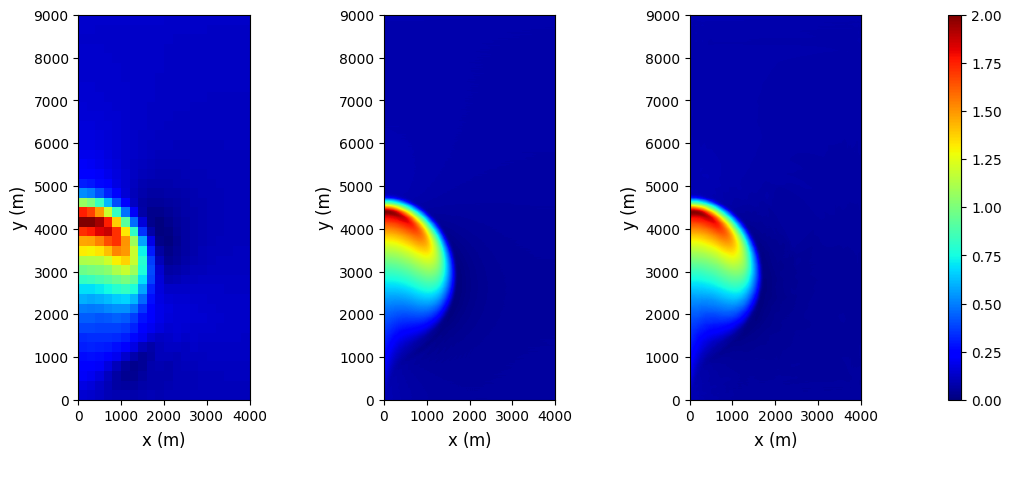}
         \put(9,40){\scriptsize{\textcolor{white}{$t/T=0.5$}}}
        \put(38,40){\scriptsize{\textcolor{white}{$t/T=0.5$}}}
        \put(68,40){\scriptsize{\textcolor{white}{$t/T=0.5$}}}
                \put(13,47){\footnotesize{$I_{LR}$}}
        \put(41,47){\footnotesize{$I_{HR}$}}
        \put(63,47){\footnotesize{SR with \emph{CNN}}}
        \put(91,47){\footnotesize{$\theta'$}}
    \end{overpic}\\
    \vskip .2cm
\begin{overpic}[width=0.49\textwidth,grid=false]{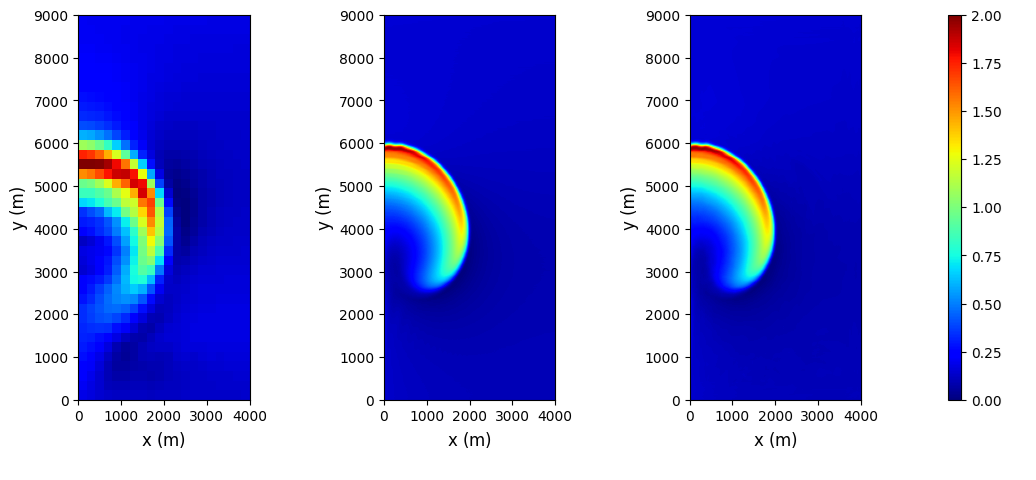}
        \put(8,40){\scriptsize{\textcolor{white}{$t/T=0.75$}}}
        \put(37.5,40){\scriptsize{\textcolor{white}{$t/T=0.75$}}}
        \put(67.5,40){\scriptsize{\textcolor{white}{$t/T=0.75$}}}
                \put(13,47){\footnotesize{$I_{LR}$}}
        \put(41,47){\footnotesize{$I_{HR}$}}
        \put(63,47){\footnotesize{SR with \emph{CNN}}}
        \put(91,47){\footnotesize{$\theta'$}}
    \end{overpic} 
\begin{overpic}[width=0.49\textwidth,grid=false]{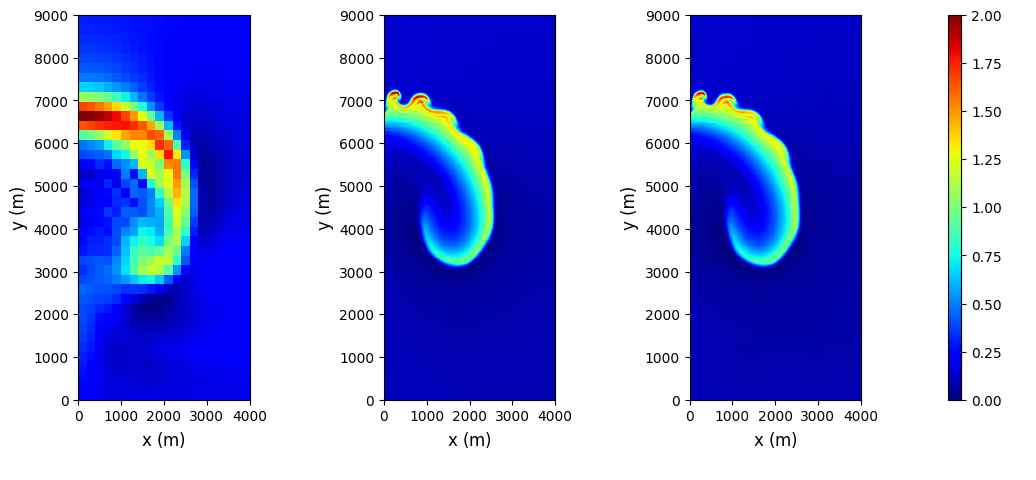}
        \put(10,40){\scriptsize{\textcolor{white}{$t/T=1$}}}
        \put(40,40){\scriptsize{\textcolor{white}{$t/T=1$}}}
         \put(70,40){\scriptsize{\textcolor{white}{$t/T=1$}}}
        \put(13,47){\footnotesize{$I_{LR}$}}
        \put(41,47){\footnotesize{$I_{HR}$}}
        \put(63,47){\footnotesize{SR with \emph{CNN}}}
        \put(91,47){\footnotesize{$\theta'$}}
    \end{overpic}
    
    \caption{Rising bubble, coarse mesh 
    $h = 250$ m: low-resolution solution computed  by the AV model (left in each panel), reference solution computed by the Smagorinsky model (center in each panel), and improvement by the super resolution with \emph{CNN} (right in each panel) for different times $t$, with $T = 1020$ s.}
    \label{fig:250_32_LES}
\end{figure}

\begin{figure}[htb!]
    \centering
\begin{overpic}[width=0.49\textwidth,grid=false]{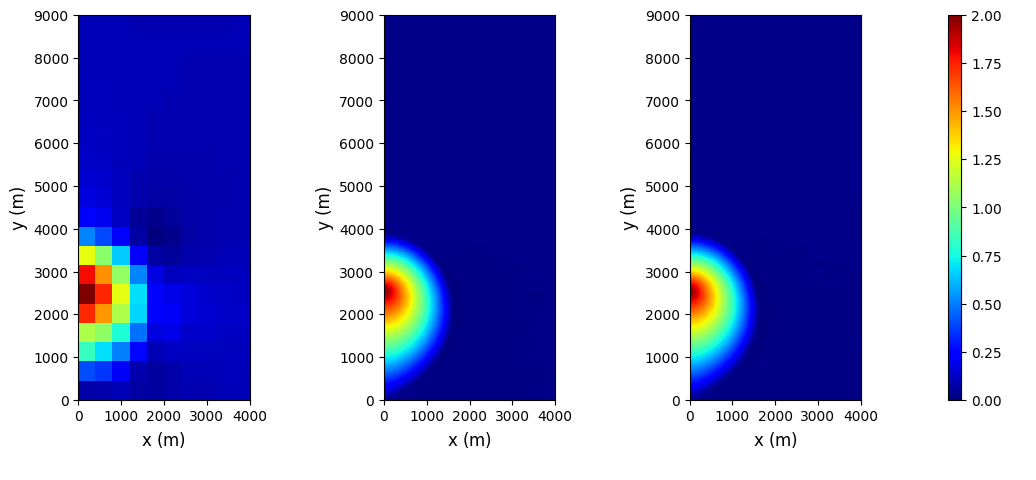}
        \put(8,40){\scriptsize{\textcolor{white}{$t/T=0.25$}}}
        \put(38,40){\scriptsize{\textcolor{white}{$t/T=0.25$}}}
        \put(67,40){\scriptsize{\textcolor{white}{$t/T=0.25$}}}
        \put(13,47){\footnotesize{$I_{LR}$}}
        \put(41,47){\footnotesize{$I_{HR}$}}
        \put(63,47){\footnotesize{SR with \emph{CNN}}}
        \put(91,47){\footnotesize{$\theta'$}}
    \end{overpic} 
\begin{overpic}[width=0.49\textwidth,grid=false]{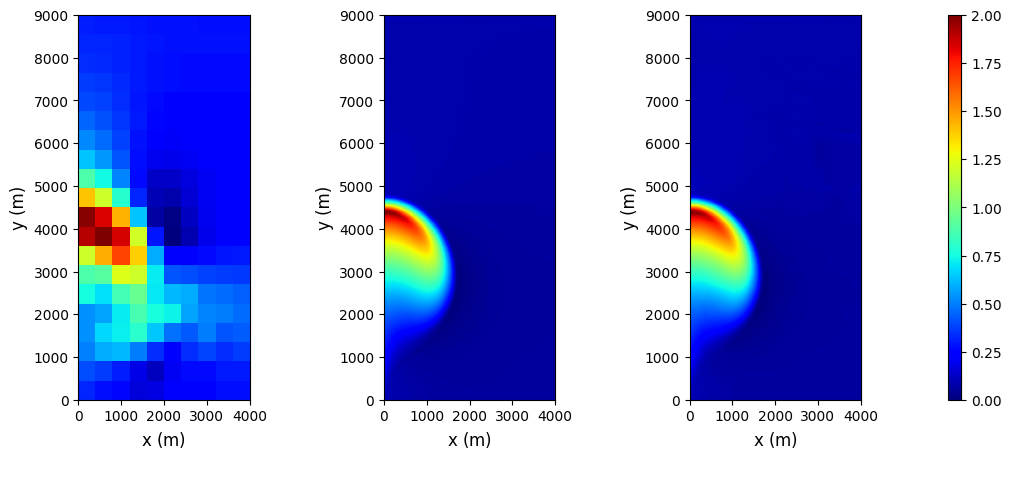}
         \put(9,40){\scriptsize{\textcolor{white}{$t/T=0.5$}}}
        \put(38,40){\scriptsize{\textcolor{white}{$t/T=0.5$}}}
        \put(68,40){\scriptsize{\textcolor{white}{$t/T=0.5$}}}
                \put(13,47){\footnotesize{$I_{LR}$}}
        \put(41,47){\footnotesize{$I_{HR}$}}
        \put(63,47){\footnotesize{SR with \emph{CNN}}}
        \put(91,47){\footnotesize{$\theta'$}}
    \end{overpic}\\
    \vskip .2cm
\begin{overpic}[width=0.49\textwidth,grid=false]{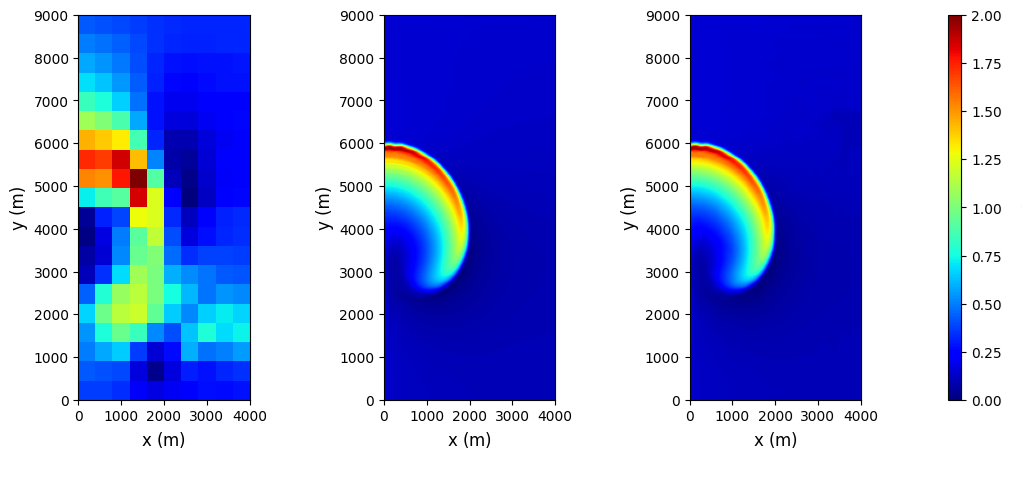}
        \put(8,40){\scriptsize{\textcolor{white}{$t/T=0.75$}}}
        \put(37.5,40){\scriptsize{\textcolor{white}{$t/T=0.75$}}}
        \put(67.5,40){\scriptsize{\textcolor{white}{$t/T=0.75$}}}
                \put(13,47){\footnotesize{$I_{LR}$}}
        \put(41,47){\footnotesize{$I_{HR}$}}
        \put(63,47){\footnotesize{SR with \emph{CNN}}}
        \put(91,47){\footnotesize{$\theta'$}}
    \end{overpic} 
\begin{overpic}[width=0.49\textwidth,grid=false]{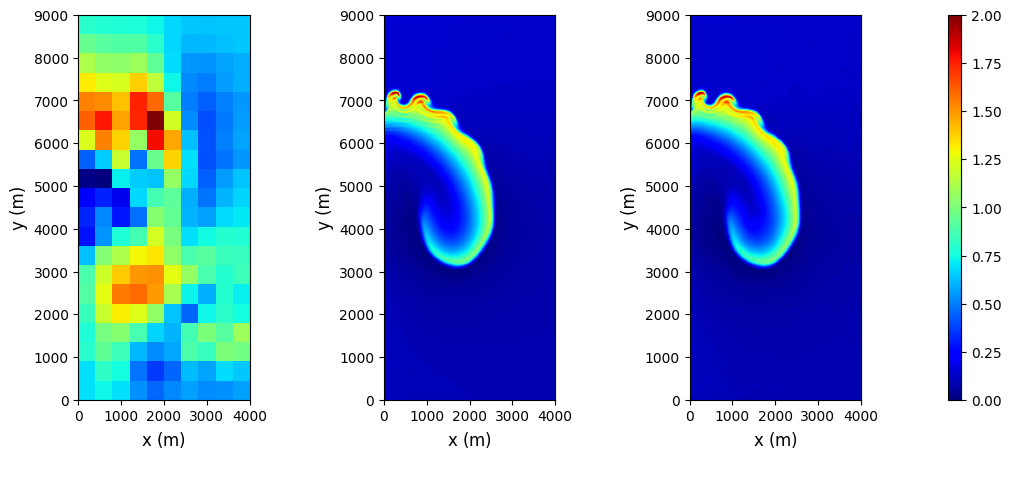}
        \put(10,40){\scriptsize{{$t/T=1$}}}
        \put(40,40){\scriptsize{\textcolor{white}{$t/T=1$}}}
         \put(70,40){\scriptsize{\textcolor{white}{$t/T=1$}}}
        \put(13,47){\footnotesize{$I_{LR}$}}
        \put(41,47){\footnotesize{$I_{HR}$}}
        \put(63,47){\footnotesize{SR with \emph{CNN}}}
        \put(91,47){\footnotesize{$\theta'$}}
    \end{overpic}
    
    \caption{Rising bubble, coarse mesh 
    $h = 500$ m: low resolution solution computed by the AV model (left in each panel), reference solution computed by the Smagorinsky model (center in each panel), and improvement by the super resolution with \emph{CNN} (right in each panel) for different times $t$, with $T = 1020$ s.}
    \label{fig:500_32_LES}
\end{figure}

Next, we consider as high-resolution dataset  $I_{HR}$ the solutions provided by the Smagorinsky model with a structured mesh with grid size $h = 32.5$ m 
and time $\Delta t = 0.0125$ s. Note that for the reference solution we are using a finer mesh than in the case of the AV model so that we can observe a well-developed Rayleigh-Taylor
instability at the top of the bubble. 
See, e.g., \citep{GQR_OF_clima, GirfoglioFVCA10, GIRFOGLIO2025106510}, for more comparisons on the solutions provided by the Smagorinsky model and the AV model for a given mesh. 
The low-resolution data $I_{LR}$ are the solutions computed by the AV model 
with four coarser structured meshes with mesh size $h=62.5, 125, 250, 500$ m and time step $\Delta t = 0.0125$ s. 
Fig.~\ref{fig:62_32_LES}-\ref{fig:500_32_LES} 
compare the reference solutions in $I_{HR}$
with the solutions 
obtained with the coarse meshes and their improvement through SR with the standard $\emph{CNN}$.
We see that the low-resolution solutions 
do not capture
the Rayleigh-Taylor instability at the top of the bubble with the coarser meshes, but the instability emerges when the such solutions are enhanced
by SR with the
\emph{CNN} architecture. 
Overall, we see great qualitative agreement between the low-resolution solutions are enhanced by SR and the reference (high-resolution) solutions. 

\begin{figure}[htb]
    \centering
\begin{overpic}[width=0.49\textwidth,grid=false]{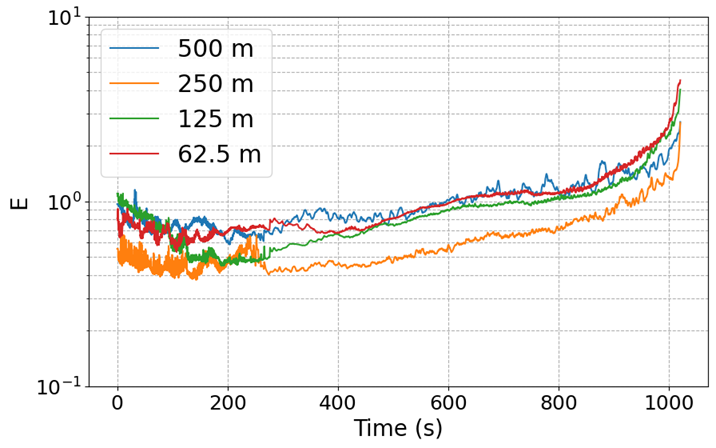}
    \end{overpic}
    \caption{Rising bubble, Smagorinsky model for the reference solution: evolution of error \eqref{eq:L2_norm} for the solutions computed by the AV model with coarse meshes  $h = 62.5, 125, 250, 500$ m.}
    \label{fig:L2_norm_S}
\end{figure}

For a quantitative comparison of the plots in 
Fig.~\ref{fig:62_32_LES}-\ref{fig:500_32_LES},
Fig.~\ref{fig:L2_norm_S} shows the evolution of error (\ref{eq:L2_norm}). 
We observe that, while in Fig.~\ref{fig:L2_norm_AV}
the error oscillated around roughly 0.8\% and did not exceed 2\%, in Fig.~\ref{fig:L2_norm_S} the error steadily increases with time starting from roughly $t = 300$ s and it reaches 4\% at the end of the simulation. 
We also note that for the errors in Fig.~\ref{fig:L2_norm_AV} the low-resolution and high-resolution solutions were obtained with the same model (AV model), while for the 
errors in Fig.~\ref{fig:L2_norm_S} the model is different. This might play a role in the increase of the errors. In addition, fewer snapshots were used for training when the high-resolution dataset comes from the Smagorinsky model due to a trade-off between accuracy and computational efficiency. More snapshots could improve the errors in 
Fig.~\ref{fig:L2_norm_S} at the price of an increased computational cost. Finally, 
Fig.~\ref{fig:L2_norm_S} might indicate that the super-resolution with the \emph{CNN} architecture reaches the limits of its applicability
when the flow becomes more complex and the time intervals of interest are long.

\begin{figure}[htb!]
     \centering
     
     \begin{subfigure}[b]{0.23\textwidth}
         \centering
         \begin{overpic}[width=0.8\textwidth]{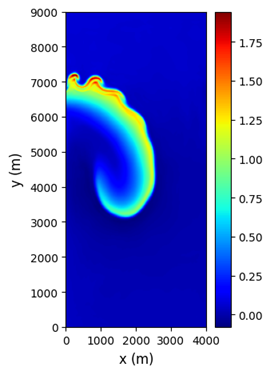}
            \put(73,50){\color{black}$\theta'$} 
            \put(27,87){\color{white}{80\%}}
         \end{overpic}
         \label{fig:80_perc-RTB}
     \end{subfigure}
     \begin{subfigure}[b]{0.23\textwidth}
         \centering
         \begin{overpic}[width=0.8\textwidth]{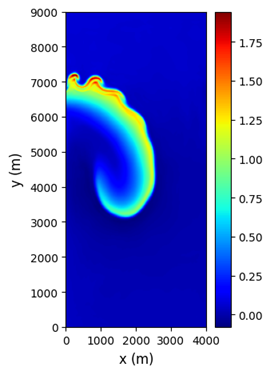}
            \put(73,50){\color{black}$\theta'$} 
            \put(27,87){\color{white}{60\%}}
         \end{overpic}
         \label{fig:60_perc-RTB}
     \end{subfigure}
     \begin{subfigure}[b]{0.23\textwidth}
         \centering
         \begin{overpic}[width=0.8\textwidth]{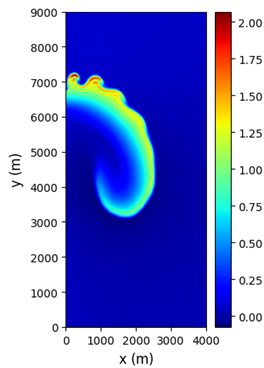}
            \put(73,50){\color{black}$\theta'$} 
            \put(27,87){\color{white}{40\%}}
         \end{overpic}
         \label{fig:40_perc-RTB}
     \end{subfigure}
     \begin{subfigure}[b]{0.23\textwidth}
         \centering
         \begin{overpic}[width=0.8\textwidth]{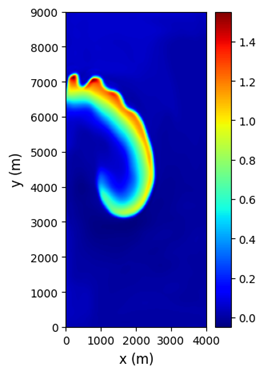}
            \put(73,50){\color{black}$\theta'$} 
            \put(27,87){\color{white}{20\%}}
         \end{overpic}
         \label{fig:20_perc-RTB}
     \end{subfigure}
     
     \caption{Rising bubble, coarse mesh 
    $h =500$~m: improvements to the low-resolution solution computed by the AV model at time $T = 1020$ s by the super resolution with \emph{CNN} trained with 
    80\%, 60\%, 40\%, 20\% of the dataset. The $I_{HR}$
    dataset consists of solutions computed by the Smagorinsky model.}
     \label{fig:RTB_comparison}
\end{figure}
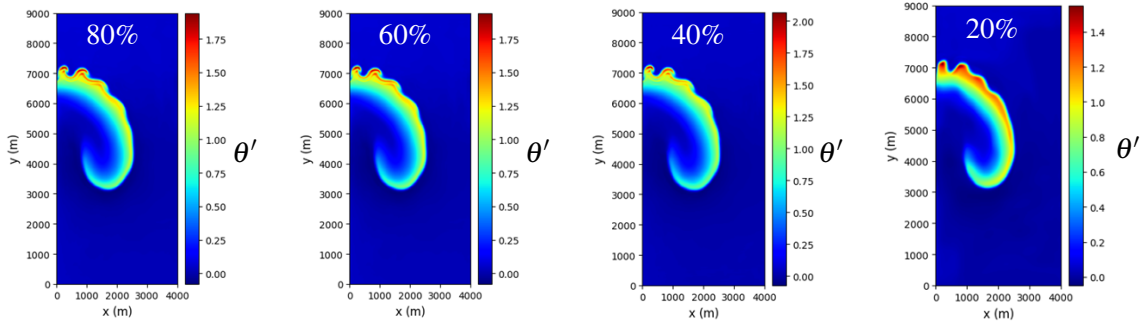

Next, we investigate the sensitivity to
the amount of data used for training. Specifically, we want to understand
the extent to which the training requirements 
can be minimized without compromising the fidelity of the reconstruction. Thus, we 
vary the training-to-testing ratio to identify the critical threshold for effective SR learning.
Fig.~\ref{fig:RTB_comparison} shows the 
improvements to the low-resolution solution computed by the AV model (mesh $h = 500 $~m) at time $T = 1020$ s by the super resolution with \emph{CNN} trained with 80\%, 60\%, 40\%, 20\% of the dataset. The dataset consists of solutions computed by the Smagorinsky model, sampled as explained above. We see that the SR with 
\emph{CNN} is remarkably robust, maintaining a
high-fidelity flow field reconstruction even 
as the training data shrinks. When the training allocation is maintained between $80\%$ and $60\%$, we obtain a nearly identical reconstruction of $\theta'$, with 
a negligible loss in accuracy.
At the $40\%$ training threshold, we observe a
degeneration in accuracy: while the shape
of the warm bubble is correctly captured, 
the magnitude of $\theta'$ is off. 
At the $20\%$ training threshold,
the reconstruction fails to capture the essential physics of the vortical structures. 
We note that for brevity Fig.~\ref{fig:RTB_comparison} shows only the results at the final time step, but similar
accuracy is seen over the entire time interval
of interest. 

\subsection{Density current}\label{sec:DC}

The computational domain for this benchmark in the $xz$-plane is $\Omega =[0,25600] \times [0,6400]$ m$^2$ and the time interval of interest is $(0, 900]$ s. 
The initial density is given by eq. \eqref{eq:rho_wb}. The initial potential temperature $\theta^0$ is defined as: 
\begin{equation}
\theta^0 = 300 - \frac{15}{2}\left[  1 + \cos(\pi r)\right] ~ \textrm{if $r\leq 1$},\quad\theta^0 = 300
~ \textrm{otherwise},
\label{warmEqn2}
\end{equation}
where $r = \sqrt[]{\left(\frac{x-x_{c}}{x_r}\right)^{2} + \left(\frac{z-z_{c}}{z_r}\right)^{2}}$, with $(x_r,z_r)=(4000, 2000)~{\rm m}$ and $(x_c,z_c) = (0,3000)~\mathrm{m}$ \cite{clinco2023filter}. 
In this benchmark, the perturbation is a circular bubble of colder air. The initial velocity field is zero everywhere and the initial
total energy is given by \eqref{eq:e0}. We impose impenetrable, free-slip boundary conditions on all the boundary. 


In this section, we will first show that the SR with standard \emph{CNN} fails at this more complex benchmark and then switch to the improved architectures (\emph{A-CNN}, \emph{m-CNN}, and \emph{Diff}). The hyperparameters for the \emph{A-CNN}, \emph{m-CNN} and \emph{Diff} networks are found in Tab.~\ref{tab:cnn_att_hyperparameters}, \ref{tab:cnn_ms_hyperparameters} and \ref{tab:diffusion_cnn_hyperparameters}. 
We will start by considering reference solutions from the AV model with a
structured mesh with grid size $h=100$ m 
and time step $\Delta t = 0.1$ s. We sample the simulation $9000$ times (i.e., every time step). Again, we start with 
80\% of the snapshots used for training, while the remaining 20\% are reserved for validation. We will switch to reference
solutions given by the Smagorinsky model
in Sec.~\ref{sec:tradeoff}, where we will also vary the split the percentage of the dataset used for training.

\begin{figure}[htb!]
   \centering
    %
        
\begin{overpic}[width=0.49\textwidth,grid=false]{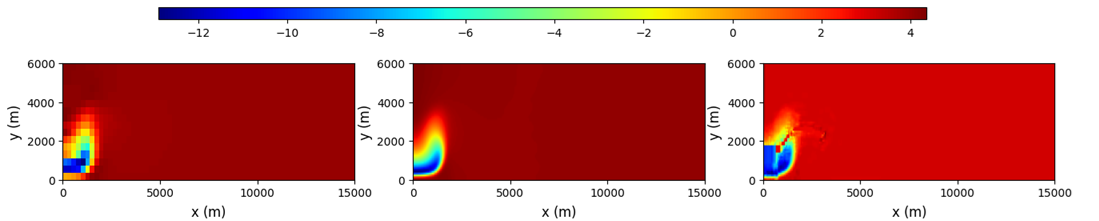}
        \put(11,11){\scriptsize{\textcolor{white}{$t/T=0.25$}}}
        \put(45,11){\scriptsize{\textcolor{white}{$t/T=0.25$}}}
        \put(77,11){\scriptsize{\textcolor{white}{$t/T=0.25$}}}
        \put(15,-3){\footnotesize{$I_{LR}$}}
        \put(46,-3){\footnotesize{$I_{HR}$}}
        \put(73,-3){\footnotesize{SR with \emph{CNN}}}
        \put(88,18){\footnotesize{$\theta'$}}
        \put(46,22){\footnotesize{$h =400$}}
    \end{overpic} 
    \begin{overpic}[width=0.49\textwidth,grid=false]{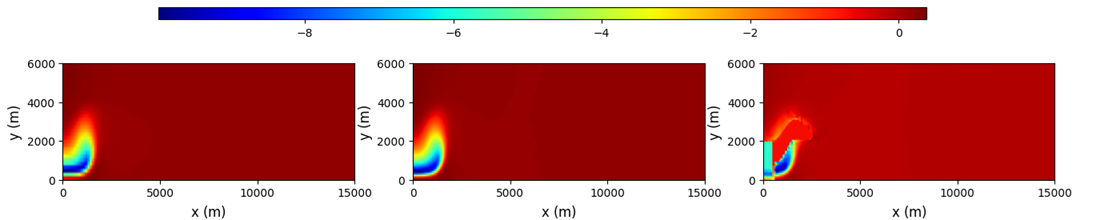}
        \put(11,11){\scriptsize{\textcolor{white}{$t/T=0.25$}}}
        \put(45,11){\scriptsize{\textcolor{white}{$t/T=0.25$}}}
        \put(77,11){\scriptsize{\textcolor{white}{$t/T=0.25$}}}
        \put(15,-3){\footnotesize{$I_{LR}$}}
        \put(47,-3){\footnotesize{$I_{HR}$}}
        \put(75,-3){\footnotesize{SR with \emph{CNN}}}
        \put(88,18){\footnotesize{$\theta'$}}
        \put(46,22){\footnotesize{$h =200$}}
    \end{overpic}
    \\
    \vskip .4cm
    \begin{overpic}[width=0.49\textwidth,grid=false]{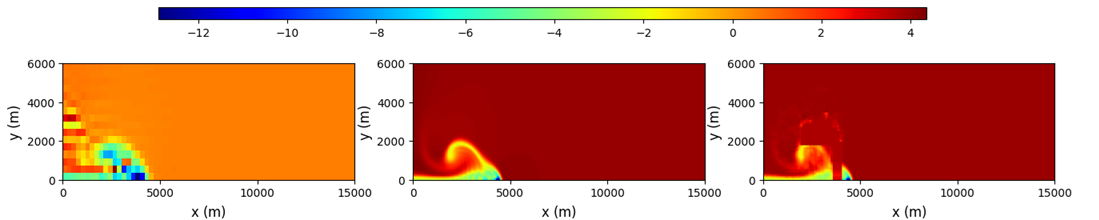}
        \put(11,11){\scriptsize{\textcolor{white}{$t/T=0.5$}}}
        \put(45,11){\scriptsize{\textcolor{white}{$t/T=0.5$}}}
        \put(78,11){\scriptsize{\textcolor{white}{$t/T=0.5$}}}
        \put(15,-3){\footnotesize{$I_{LR}$}}
        \put(46,-3){\footnotesize{$I_{HR}$}}
        \put(73,-3){\footnotesize{SR with \emph{CNN}}}
        \put(88,18){\footnotesize{$\theta'$}}
    \end{overpic} 
        \begin{overpic}[width=0.49\textwidth,grid=false]{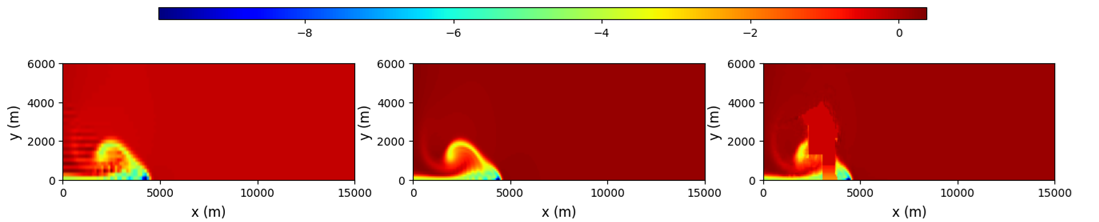}
        \put(11,11){\scriptsize{\textcolor{white}{$t/T=0.5$}}}
        \put(45,11){\scriptsize{\textcolor{white}{$t/T=0.5$}}}
        \put(78,11){\scriptsize{\textcolor{white}{$t/T=0.5$}}}
        \put(17,-3){\footnotesize{$I_{LR}$}}
        \put(47,-3){\footnotesize{$I_{HR}$}}
        \put(75,-3){\footnotesize{SR with \emph{CNN}}}
        \put(88,18){\footnotesize{$\theta'$}}
    \end{overpic}
    \\
    \vskip .5cm
\begin{overpic}[width=0.49\textwidth,grid=false]{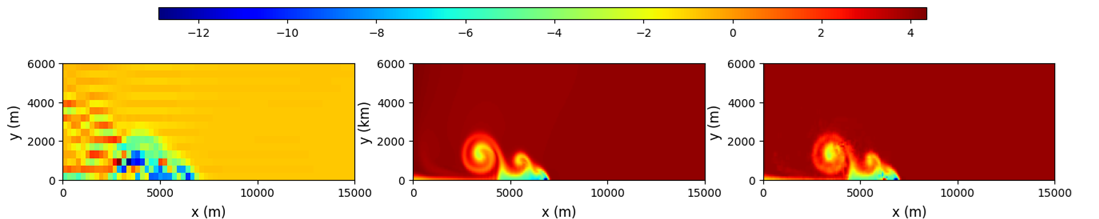}
        \put(11,11){\scriptsize{\textcolor{black}{$t/T=0.75$}}}
        \put(45,11){\scriptsize{\textcolor{white}{$t/T=0.75$}}}
        \put(77,11){\scriptsize{\textcolor{white}{$t/T=0.75$}}}
       \put(15,-3){\footnotesize{$I_{LR}$}}
        \put(46,-3){\footnotesize{$I_{HR}$}}
        \put(73,-3){\footnotesize{SR with \emph{CNN}}}
        \put(88,18){\footnotesize{$\theta'$}}
    \end{overpic}
    \begin{overpic}[width=0.49\textwidth,grid=false]{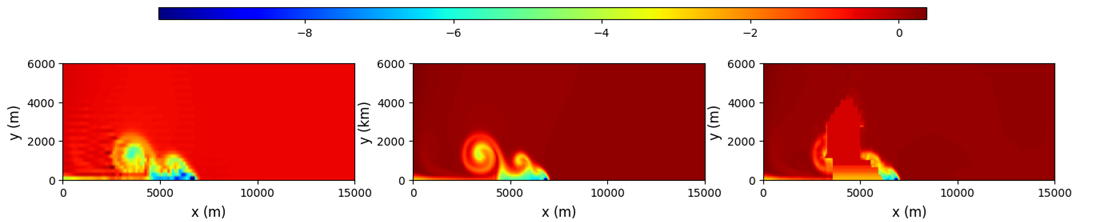}
        \put(11,11){\scriptsize{\textcolor{white}{$t/T=0.75$}}}
        \put(45,11){\scriptsize{\textcolor{white}{$t/T=0.75$}}}
        \put(77,11){\scriptsize{\textcolor{white}{$t/T=0.75$}}}
        \put(15,-3){\footnotesize{$I_{LR}$}}
        \put(47,-3){\footnotesize{$I_{HR}$}}
        \put(75,-3){\footnotesize{SR with \emph{CNN}}}
        \put(88,18){\footnotesize{$\theta'$}}
    \end{overpic}
    \\
    \vskip .4cm
    \begin{overpic}[width=0.49\textwidth,grid=false]{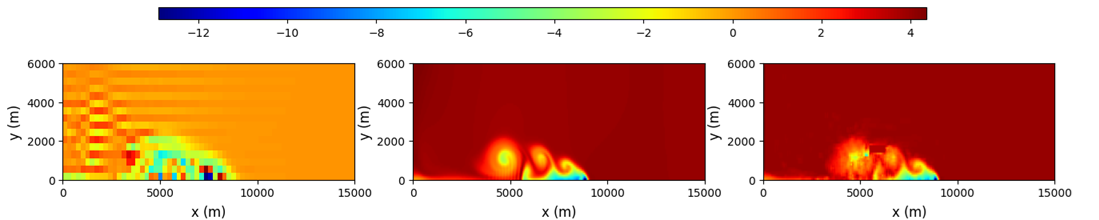}
        \put(11,11){\scriptsize{\textcolor{white}{$t/T=1$}}}
        \put(45,11){\scriptsize{\textcolor{white}{$t/T=1$}}}
        \put(77,11){\scriptsize{\textcolor{white}{$t/T=1$}}}
        \put(15,-3){\footnotesize{$I_{LR}$}}
        \put(46,-3){\footnotesize{$I_{HR}$}}
        \put(73,-3){\footnotesize{SR with \emph{CNN}}}
        \put(88,18){\footnotesize{$\theta'$}}
    \end{overpic}
        \begin{overpic}[width=0.49\textwidth,grid=false]{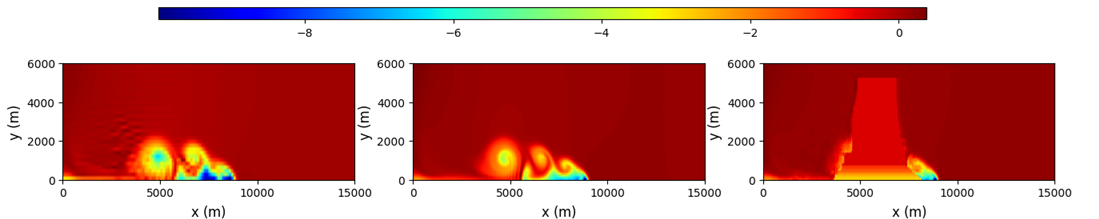}
        \put(11,11){\scriptsize{\textcolor{white}{$t/T=1$}}}
        \put(45,11){\scriptsize{\textcolor{white}{$t/T=1$}}}
        \put(77,11){\scriptsize{\textcolor{white}{$t/T=1$}}}
        \put(15,-3){\footnotesize{$I_{LR}$}}
        \put(47,-3){\footnotesize{$I_{HR}$}}
        \put(75,-3){\footnotesize{SR with \emph{CNN}}}
        \put(88,18){\footnotesize{$\theta'$}}
    \end{overpic}
    \caption{Density current, coarse mesh 
    $h = 400$ m (left panel) and $h = 200$ m (right panel): low resolution solution computed by the AV model ($I_{LR}$), reference solution computed by the AV model ($I_{HR}$), and improvement by the super resolution with \emph{CNN} (``SR with \emph{CNN}'') for increasing time $t$ from top to bottom, with $T = 900$ s.}
    \label{fig:400_100_classic}
\end{figure}

The low resolution data $I_{LR}$ are the solutions computed by the AV model with two coarse structured meshes with mesh size $h=200, 400$ m
and time step $\Delta t = 0.1$ s.
Fig. \ref{fig:400_100_classic} 
compares the reference solutions in $I_{HR}$ with the solutions 
obtained with the coarse meshes and their improvement through the standard $CNN$ architecture.
Due to the presence of more complex 
flow structures, we see that the \emph{CNN}
architecture that worked well for the 
rising thermal bubble benchmark (see Sec.~\ref{sec:RTB}) fails for the density current benchmark. 

\begin{figure}[htb!]
   \centering
\begin{overpic}[width=0.9\textwidth,grid=false]{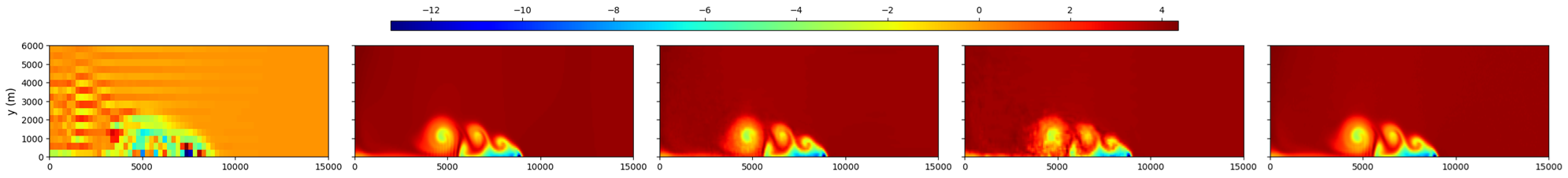}
        \put(6,6){\scriptsize{\textcolor{white}{$t/T=0.25$}}}
        \put(26,6){\scriptsize{\textcolor{white}{$t/T=0.25$}}}
        \put(43,6){\scriptsize{\textcolor{white}{$t/T=0.25$}}}
        \put(63,6){\scriptsize{\textcolor{white}{$t/T=0.25$}}}
        \put(83,6){\scriptsize{\textcolor{white}{$t/T=0.25$}}}
        \put(10,-2){\footnotesize{$I_{LR}$}}
        \put(30,-2){\footnotesize{$I_{HR}$}}
        \put(43,-2){\footnotesize{SR with \emph{m-CNN}}}
        \put(63,-2){\footnotesize{SR with \emph{A-CNN}}}
        \put(85,-2){\footnotesize{SR with \emph{Diff}}}
        \put(78,10){\footnotesize{$\theta'$}}
    \end{overpic} \\
    \vskip .5cm
    \begin{overpic}[width=0.9\textwidth,grid=false]{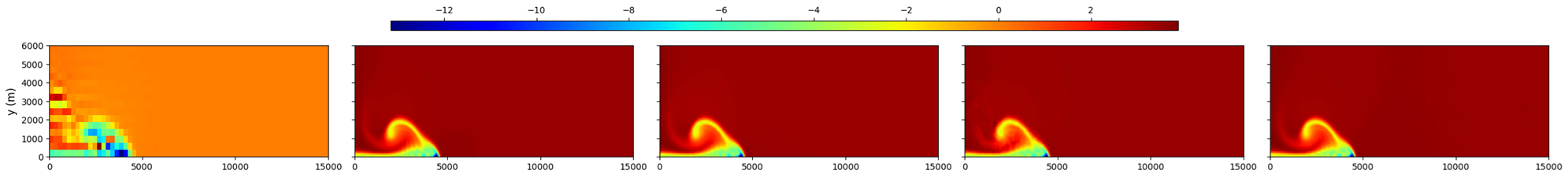}
        \put(6,6){\scriptsize{\textcolor{white}{$t/T=0.5$}}}
        \put(26,6){\scriptsize{\textcolor{white}{$t/T=0.5$}}}
        \put(43,6){\scriptsize{\textcolor{white}{$t/T=0.5$}}}
        \put(63,6){\scriptsize{\textcolor{white}{$t/T=0.5$}}}
        \put(83,6){\scriptsize{\textcolor{white}{$t/T=0.5$}}}
        \put(10,-2){\footnotesize{$I_{LR}$}}
        \put(30,-2){\footnotesize{$I_{HR}$}}
        \put(43,-2){\footnotesize{SR with \emph{m-CNN}}}
        \put(63,-2){\footnotesize{SR with \emph{A-CNN}}}
        \put(85,-2){\footnotesize{SR with \emph{Diff}}}
        \put(78,10){\footnotesize{$\theta'$}}
    \end{overpic} \\
    \vskip .5cm
\begin{overpic}[width=0.9\textwidth,grid=false]{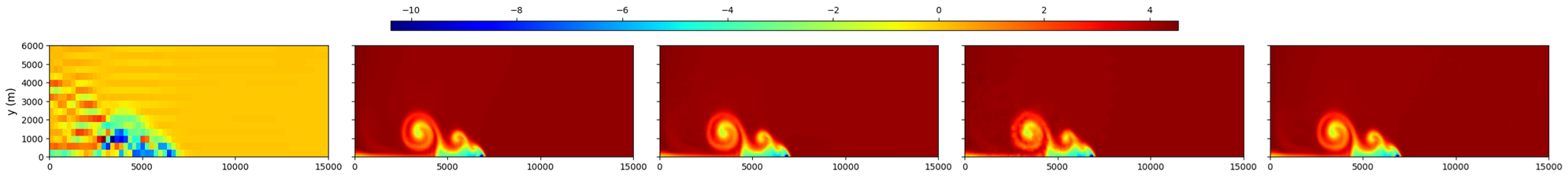}
        \put(6,6){\scriptsize{\textcolor{white}{$t/T=0.75$}}}
        \put(26,6){\scriptsize{\textcolor{white}{$t/T=0.75$}}}
        \put(43,6){\scriptsize{\textcolor{white}{$t/T=0.75$}}}
        \put(63,6){\scriptsize{\textcolor{white}{$t/T=0.75$}}}
        \put(83,6){\scriptsize{\textcolor{white}{$t/T=0.75$}}}
        \put(10,-2){\footnotesize{$I_{LR}$}}
        \put(30,-2){\footnotesize{$I_{HR}$}}
        \put(43,-2){\footnotesize{SR with \emph{m-CNN}}}
        \put(63,-2){\footnotesize{SR with \emph{A-CNN}}}
        \put(85,-2){\footnotesize{SR with \emph{Diff}}}
        \put(78,10){\footnotesize{$\theta'$}}
    \end{overpic}\\
    \vskip .5cm
    \begin{overpic}[width=0.9\textwidth,grid=false]{Figures/diff_SR_400_100_LES_t_1_2.png}
       \put(6,6){\scriptsize{\textcolor{white}{$t/T=1$}}}
        \put(26,6){\scriptsize{\textcolor{white}{$t/T=1$}}}
        \put(43,6){\scriptsize{\textcolor{white}{$t/T=1$}}}
        \put(63,6){\scriptsize{\textcolor{white}{$t/T=1$}}}
        \put(83,6){\scriptsize{\textcolor{white}{$t/T=1$}}}
        \put(10,-2){\footnotesize{$I_{LR}$}}
        \put(30,-2){\footnotesize{$I_{HR}$}}
        \put(43,-2){\footnotesize{SR with \emph{m-CNN}}}
        \put(63,-2){\footnotesize{SR with \emph{A-CNN}}}
        \put(85,-2){\footnotesize{SR with \emph{Diff}}}
        \put(78,10){\footnotesize{$\theta'$}}
    \end{overpic}
    
    \caption{Density current, coarse mesh $h = 400$ m: low resolution solution computed by the AV model (first column), reference solution computed by the AV model (second column), improvement by the super resolution with \emph{m-CNN} (third column), \emph{A-CNN} (fourth column), and \emph{Diff} (fifth column) for increasing time $t$ from top to bottom, with $T = 900$ s.
}
    \label{fig:400_100_ms_att_comp}
\end{figure}
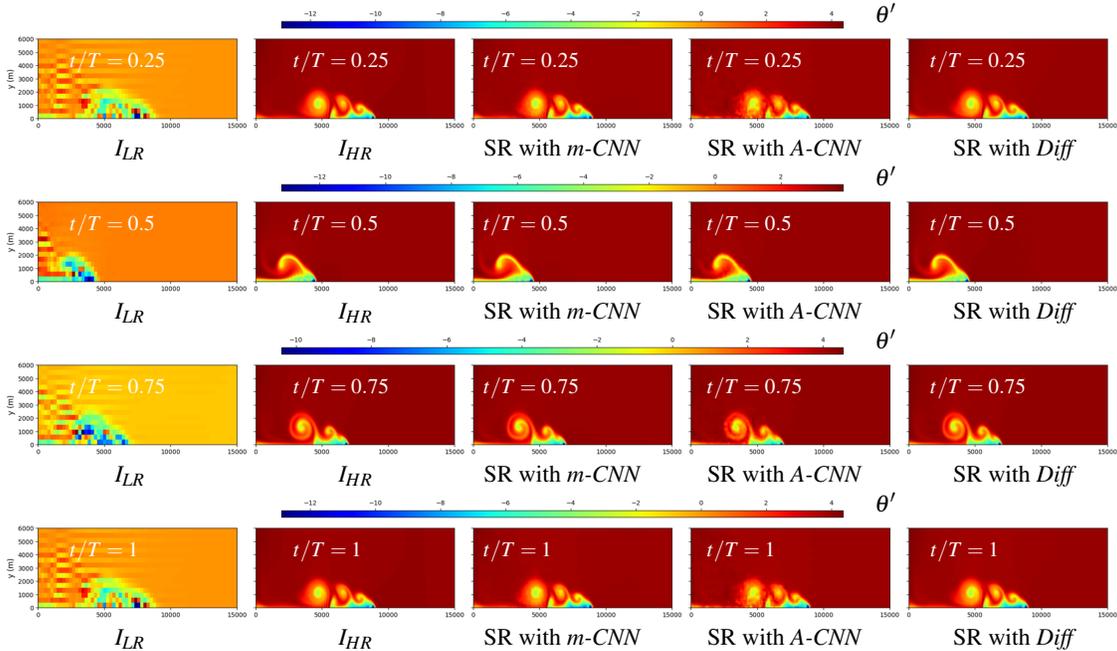

\begin{figure}[htb!]
   \centering
\begin{overpic}[width=0.9\textwidth,grid=false]{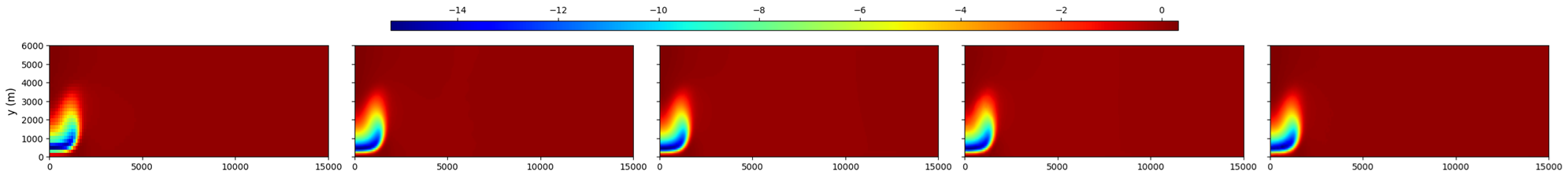}
        \put(6,6){\scriptsize{\textcolor{white}{$t/T=0.25$}}}
        \put(26,6){\scriptsize{\textcolor{white}{$t/T=0.25$}}}
        \put(43,6){\scriptsize{\textcolor{white}{$t/T=0.25$}}}
        \put(63,6){\scriptsize{\textcolor{white}{$t/T=0.25$}}}
        \put(83,6){\scriptsize{\textcolor{white}{$t/T=0.25$}}}
        \put(10,-2){\footnotesize{$I_{LR}$}}
        \put(30,-2){\footnotesize{$I_{HR}$}}
        \put(43,-2){\footnotesize{SR with \emph{m-CNN}}}
        \put(63,-2){\footnotesize{SR with \emph{A-CNN}}}
        \put(85,-2){\footnotesize{SR with \emph{Diff}}}
        \put(78,10){\footnotesize{$\theta'$}}
    \end{overpic} \\
    \vskip .5cm
    \begin{overpic}[width=0.9\textwidth,grid=false]{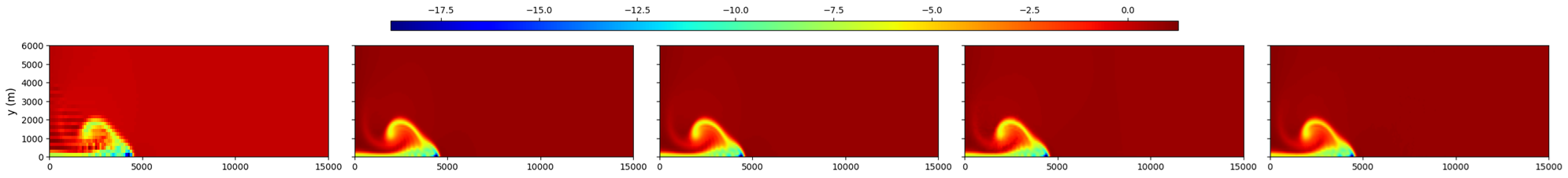}
        \put(6,6){\scriptsize{\textcolor{white}{$t/T=0.5$}}}
        \put(26,6){\scriptsize{\textcolor{white}{$t/T=0.5$}}}
        \put(43,6){\scriptsize{\textcolor{white}{$t/T=0.5$}}}
        \put(63,6){\scriptsize{\textcolor{white}{$t/T=0.5$}}}
        \put(83,6){\scriptsize{\textcolor{white}{$t/T=0.5$}}}
        \put(10,-2){\footnotesize{$I_{LR}$}}
        \put(30,-2){\footnotesize{$I_{HR}$}}
        \put(43,-2){\footnotesize{SR with \emph{m-CNN}}}
        \put(63,-2){\footnotesize{SR with \emph{A-CNN}}}
        \put(85,-2){\footnotesize{SR with \emph{Diff}}}
        \put(78,10){\footnotesize{$\theta'$}}
    \end{overpic} \\
    \vskip .5cm
\begin{overpic}[width=0.9\textwidth,grid=false]{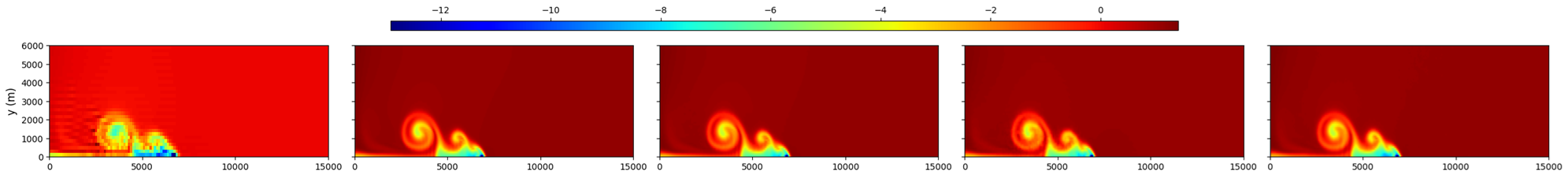}
        \put(6,6){\scriptsize{\textcolor{white}{$t/T=0.75$}}}
        \put(26,6){\scriptsize{\textcolor{white}{$t/T=0.75$}}}
        \put(43,6){\scriptsize{\textcolor{white}{$t/T=0.75$}}}
        \put(63,6){\scriptsize{\textcolor{white}{$t/T=0.75$}}}
        \put(83,6){\scriptsize{\textcolor{white}{$t/T=0.75$}}}
        \put(10,-2){\footnotesize{$I_{LR}$}}
        \put(30,-2){\footnotesize{$I_{HR}$}}
        \put(43,-2){\footnotesize{SR with \emph{m-CNN}}}
        \put(63,-2){\footnotesize{SR with \emph{A-CNN}}}
        \put(85,-2){\footnotesize{SR with \emph{Diff}}}
        \put(78,10){\footnotesize{$\theta'$}}
    \end{overpic}\\
    \vskip .5cm
    \begin{overpic}[width=0.9\textwidth,grid=false]{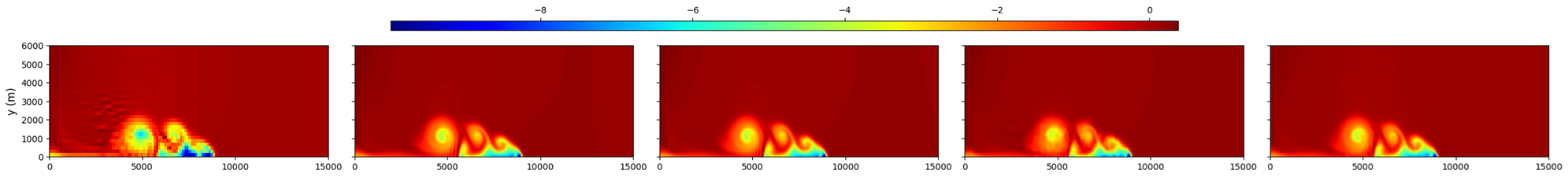}
       \put(6,6){\scriptsize{\textcolor{white}{$t/T=1$}}}
        \put(26,6){\scriptsize{\textcolor{white}{$t/T=1$}}}
        \put(43,6){\scriptsize{\textcolor{white}{$t/T=1$}}}
        \put(63,6){\scriptsize{\textcolor{white}{$t/T=1$}}}
        \put(83,6){\scriptsize{\textcolor{white}{$t/T=1$}}}
        \put(10,-2){\footnotesize{$I_{LR}$}}
        \put(30,-2){\footnotesize{$I_{HR}$}}
        \put(43,-2){\footnotesize{SR with \emph{m-CNN}}}
        \put(63,-2){\footnotesize{SR with \emph{A-CNN}}}
        \put(85,-2){\footnotesize{SR with \emph{Diff}}}
        \put(78,10){\footnotesize{$\theta'$}}
    \end{overpic}
    
    \caption{Density current, coarse mesh $h = 200$ m: low resolution solution computed by the AV model (first column), reference solution computed by the AV model (second column), improvement by the super resolution with \emph{m-CNN} (third column), \emph{A-CNN} (fourth column), and \emph{Diff} (fifth column) for increasing time $t$ from top to bottom, with $T = 900$ s.}
    \label{fig:200_100_ms_att_comp}
\end{figure}
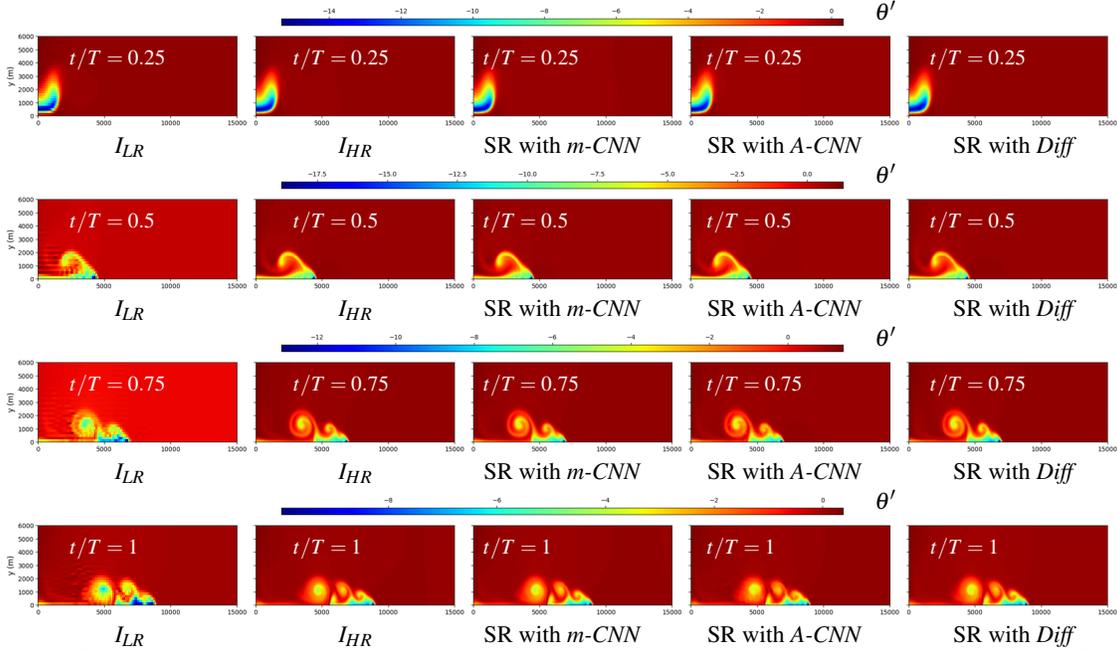

Next, we compare the reference solutions in $I_{HR}$ (same as in Fig.~\ref{fig:400_100_classic}) with the solutions obtained with
coarse meshes $h=400$ m (in Fig.~\ref{fig:400_100_ms_att_comp})
and
$h = 200$ m (in Fig.~\ref{fig:200_100_ms_att_comp})
and the improvements
by super resolution with \emph{A-CNN}, \emph{m-CNN}, and \emph{Diff}. 
Note that \emph{A-CNN} is the \emph{CNN} architecture 
used for the results in Sec.~\ref{sec:RTB} and
in Fig.~\ref{fig:400_100_classic} with the addition of an attention block. 
By comparing the fourth column in Fig.~\ref{fig:400_100_ms_att_comp} and \ref{fig:200_100_ms_att_comp}, which shows
the results obtained by SR with \emph{A-CNN},
with the third column in each panel of Fig.~\ref{fig:400_100_classic} , 
we observe a
significant improvement in accuracy thanks 
to the attention block. However, we 
see a degeneration of such improvement
towards the end of the time interval of interest. Instead, the improvements in accuracy made possible by the SR with \emph{m-CNN} (third column in Fig.~\ref{fig:400_100_ms_att_comp} and \ref{fig:200_100_ms_att_comp}) do not degenerate visibly. 
This can be explained as follows. The attention mechanism focuses on local features of an image at the pixel-level and thus it misses 
larger spatial scales, such as those observed in the density current flow. 
The multi-scale model addresses this limitation through convolutional kernels of different sizes, which can simultaneously capture small and large scales, making it a more robust tool to studying convection-dominated flows.
From Fig.~\ref{fig:400_100_ms_att_comp} and \ref{fig:200_100_ms_att_comp}, it is clear that the
multi-scale technique outperforms the attention mechanism when more complex flow structures emerge.
As for the SR with \emph{Diff}, we see it can 
successfully recover the dominant flow structures.
The visual agreement with the high-resolution solution is better than in the case
of \emph{A-CNN} and comparable to \emph{m-CNN}.

\begin{figure}[htb!]
    \centering
    \begin{minipage}{0.48\textwidth}
        \centering
        \begin{overpic}[width=\linewidth]{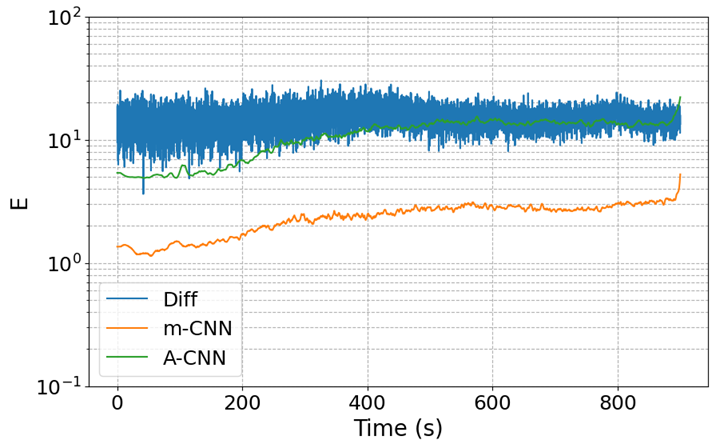}
             \put(48,-5){\small $h=400$m} 
        \end{overpic}
    \end{minipage}
    \hfill
    \begin{minipage}{0.48\textwidth}
        \centering
        \begin{overpic}[width=\linewidth]{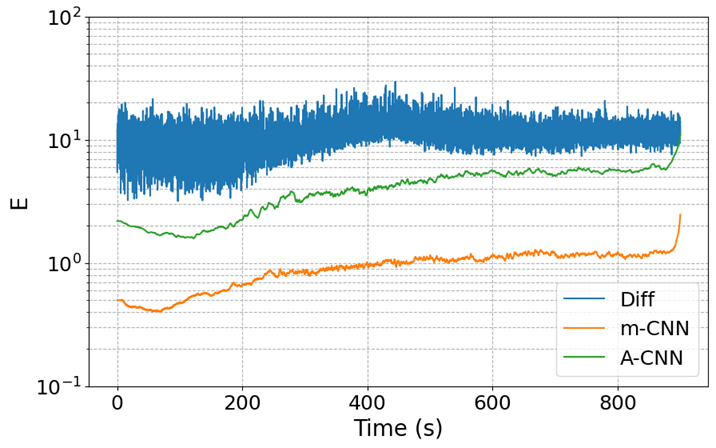}
             \put(48,-5){\small $h=200$m}
        \end{overpic}
    \end{minipage}

    \vspace{15pt} 
    \caption{Density current, AV model ( reference solution computed with mesh $h=100$): evolution of error \eqref{eq:L2_norm} 
    for coarse meshes $h = 400$ m (left) and $h = 200$ m (right) enhanced by SR with 
    \emph{A-CNN}, \emph{m-CNN}, and \emph{Diff}.
    }
    \label{fig:L2_norm_combined}
\end{figure}

For a quantitative comparison of the plots in 
Fig.~\ref{fig:400_100_ms_att_comp}-\ref{fig:200_100_ms_att_comp},
Fig. \ref{fig:L2_norm_combined} shows the evolution of error (\ref{eq:L2_norm}).
It is striking that, while the errors given by the SR with \emph{A-CNN} and \emph{m-CNN} decrease by orders of magnitude when the mesh is refined, that is not the case for the SR
with \emph{Diff}. With both meshes, the 
superior accuracy of \emph{m-CNN}
is remarkable, also because it is computationally cheaper than \emph{Diff}, which we remind
is considered the state-of-the-art for
SR tasks. 
The higher computational cost of \emph{Diff} stems from its iterative denoising process and stochastic sampling during inference. 
In contrast, the deterministic forward pass of the \emph{m-CNN} enables more efficient inference while delivering superior accuracy. Furthermore, Fig.~\ref{fig:L2_norm_combined} shows marked differences in error stability: 
the error curves for \emph{Diff} exhibit pronounced oscillations over training and evaluation alike, whereas the error curves for \emph{A-CNN} and \emph{m-CNN} have oscillations of much smaller amplitude. The large oscillations in the errors given by \emph{Diff} are likely due to the repeated injection and removal of noise intrinsic to diffusion-based training, which can introduce variance in the reconstruction quality. 


\subsubsection{Limitations and trade-offs}\label{sec:tradeoff}

As shown in Sec.~\ref{sec:RTB}, 
the solutions provided by the Smagorinsky model feature a more complex flow. When this further complexity is added to density current flows, we can more clearly see the limitations of the proposed SR approaches. In this section, we discuss such limitations and how to overcome them.

\begin{figure}[htb!]
   \centering
\begin{overpic}[width=0.8\textwidth,grid=false]{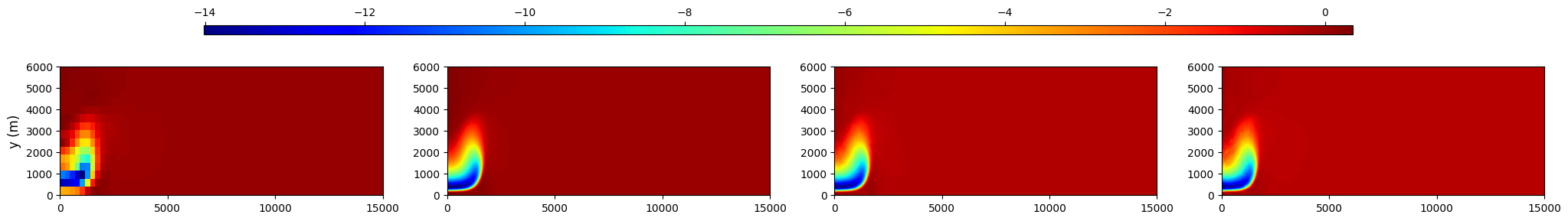}
        \put(12,7){\scriptsize{\textcolor{white}{$t/T=0.25$}}}
        \put(35,7){\scriptsize{\textcolor{white}{$t/T=0.25$}}}
        \put(60,7){\scriptsize{\textcolor{white}{$t/T=0.25$}}}
        \put(85,7){\scriptsize{\textcolor{white}{$t/T=0.25$}}}
        \put(15,-2){\footnotesize{$I_{LR}$}}
        \put(37,-2){\footnotesize{$I_{HR}$}}
        \put(55,-2){\footnotesize{SR with \emph{m-CNN}}}
        \put(80,-2){\footnotesize{SR with \emph{A-CNN}}}
        \put(88,12){\footnotesize{$\theta'$}}
    \end{overpic} \\
    \vskip .5cm
    \begin{overpic}[width=0.8\textwidth,grid=false]{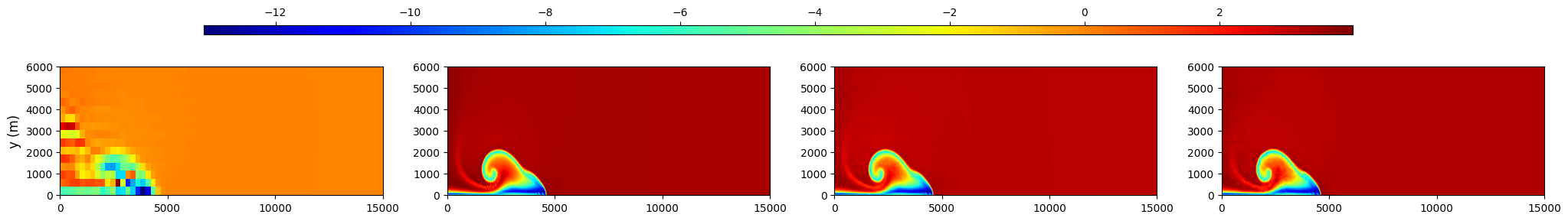}
        \put(12,7){\scriptsize{\textcolor{white}{$t/T=0.5$}}}
        \put(35,7){\scriptsize{\textcolor{white}{$t/T=0.5$}}}
        \put(60,7){\scriptsize{\textcolor{white}{$t/T=0.5$}}}
        \put(85,7){\scriptsize{\textcolor{white}{$t/T=0.5$}}}
        \put(15,-2){\footnotesize{$I_{LR}$}}
        \put(37,-2){\footnotesize{$I_{HR}$}}
        \put(55,-2){\footnotesize{SR with \emph{m-CNN}}}
        \put(80,-2){\footnotesize{SR with \emph{A-CNN}}}
        \put(88,12){\footnotesize{$\theta'$}}
    \end{overpic} \\
    \vskip .5cm
\begin{overpic}[width=0.8\textwidth,grid=false]{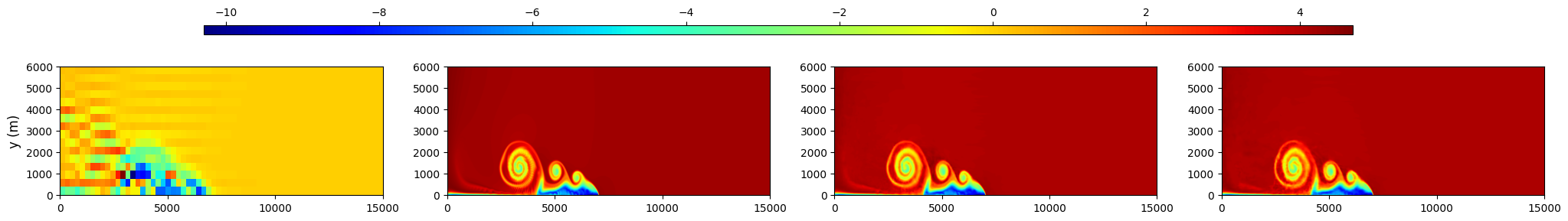}
        \put(12,7){\scriptsize{\textcolor{white}{$t/T=0.75$}}}
        \put(35,7){\scriptsize{\textcolor{white}{$t/T=0.75$}}}
        \put(60,7){\scriptsize{\textcolor{white}{$t/T=0.75$}}}
        \put(85,7){\scriptsize{\textcolor{white}{$t/T=0.75$}}}
        \put(15,-2){\footnotesize{$I_{LR}$}}
        \put(37,-2){\footnotesize{$I_{HR}$}}
        \put(55,-2){\footnotesize{SR with \emph{m-CNN}}}
        \put(80,-2){\footnotesize{SR with \emph{A-CNN}}}
        \put(88,12){\footnotesize{$\theta'$}}
    \end{overpic}\\
    \vskip .5cm
    \begin{overpic}[width=0.8\textwidth,grid=false]{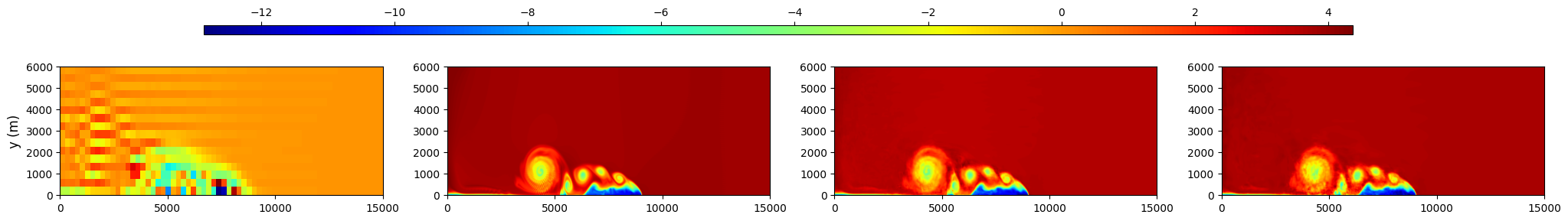}
        \put(12,7){\scriptsize{\textcolor{white}{$t/T=1$}}}
        \put(35,7){\scriptsize{\textcolor{white}{$t/T=1$}}}
        \put(60,7){\scriptsize{\textcolor{white}{$t/T=1$}}}
        \put(85,7){\scriptsize{\textcolor{white}{$t/T=1$}}}
        \put(15,-2){\footnotesize{$I_{LR}$}}
        \put(37,-2){\footnotesize{$I_{HR}$}}
        \put(55,-2){\footnotesize{SR with \emph{m-CNN}}}
        \put(80,-2){\footnotesize{SR with \emph{A-CNN}}}
        \put(88,12){\footnotesize{$\theta'$}}
    \end{overpic}
    
    \caption{Density current, coarse mesh $h = 400$ m: low resolution solution computed by the AV model (first column), reference solution computed by the Smagorinsky model (second column), improvement by the super resolution with \emph{m-CNN} (third column) and \emph{A-CNN} (fourth column) for increasing time $t$ from top to bottom, with $T = 900$ s.}
    \label{fig:400_25_ms_att_comp}
\end{figure}

\begin{figure}[htb!]
   \centering
\begin{overpic}[width=0.8\textwidth,grid=false]{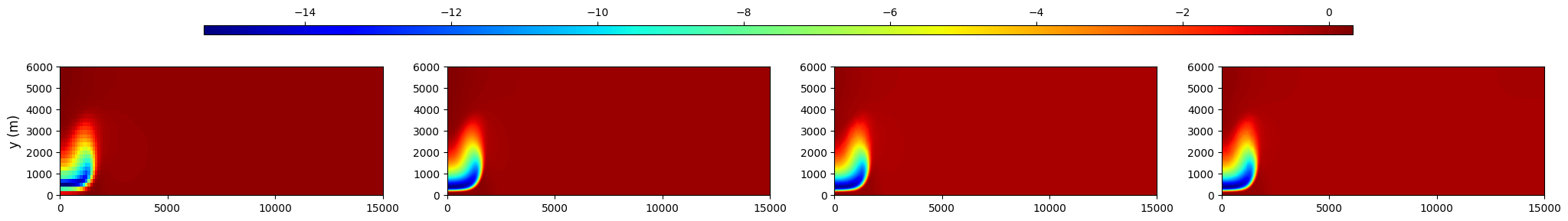}
        \put(12,7){\scriptsize{\textcolor{white}{$t/T=0.25$}}}
        \put(35,7){\scriptsize{\textcolor{white}{$t/T=0.25$}}}
        \put(60,7){\scriptsize{\textcolor{white}{$t/T=0.25$}}}
        \put(85,7){\scriptsize{\textcolor{white}{$t/T=0.25$}}}
        \put(15,-2){\footnotesize{$I_{LR}$}}
        \put(37,-2){\footnotesize{$I_{HR}$}}
        \put(55,-2){\footnotesize{SR with \emph{m-CNN}}}
        \put(80,-2){\footnotesize{SR with \emph{A-CNN}}}
        \put(88,12){\footnotesize{$\theta'$}}
    \end{overpic} \\
    \vskip .5cm
    \begin{overpic}[width=0.8\textwidth,grid=false]{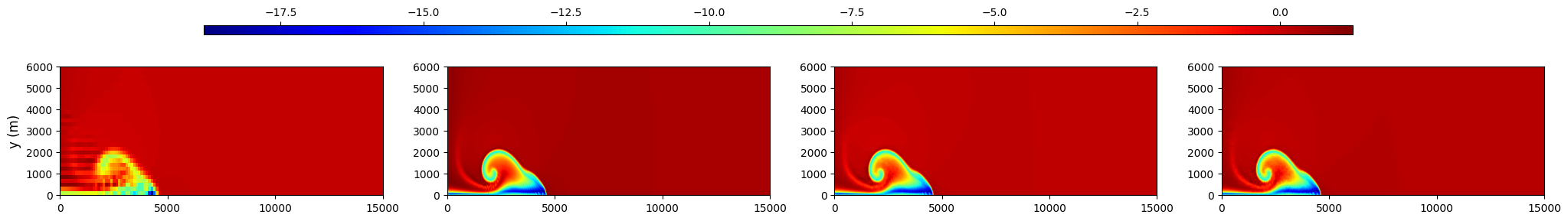}
        \put(12,7){\scriptsize{\textcolor{white}{$t/T=0.5$}}}
        \put(35,7){\scriptsize{\textcolor{white}{$t/T=0.5$}}}
        \put(60,7){\scriptsize{\textcolor{white}{$t/T=0.5$}}}
        \put(85,7){\scriptsize{\textcolor{white}{$t/T=0.5$}}}
        \put(15,-2){\footnotesize{$I_{LR}$}}
        \put(37,-2){\footnotesize{$I_{HR}$}}
        \put(55,-2){\footnotesize{SR with \emph{m-CNN}}}
        \put(80,-2){\footnotesize{SR with \emph{A-CNN}}}
        \put(88,12){\footnotesize{$\theta'$}}
    \end{overpic} \\
    \vskip .5cm
\begin{overpic}[width=0.8\textwidth,grid=false]{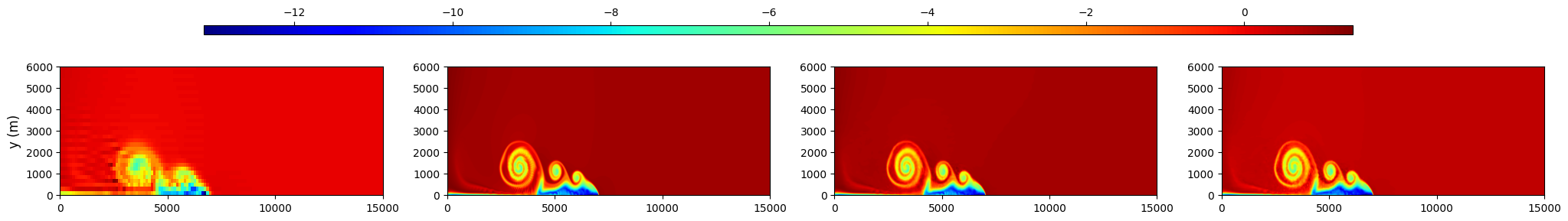}
        \put(12,7){\scriptsize{\textcolor{white}{$t/T=0.75$}}}
        \put(35,7){\scriptsize{\textcolor{white}{$t/T=0.75$}}}
        \put(60,7){\scriptsize{\textcolor{white}{$t/T=0.75$}}}
        \put(85,7){\scriptsize{\textcolor{white}{$t/T=0.75$}}}
        \put(15,-2){\footnotesize{$I_{LR}$}}
        \put(37,-2){\footnotesize{$I_{HR}$}}
        \put(55,-2){\footnotesize{SR with \emph{m-CNN}}}
        \put(80,-2){\footnotesize{SR with \emph{A-CNN}}}
        \put(88,12){\footnotesize{$\theta'$}}
    \end{overpic}\\
    \vskip .5cm
    \begin{overpic}[width=0.8\textwidth,grid=false]{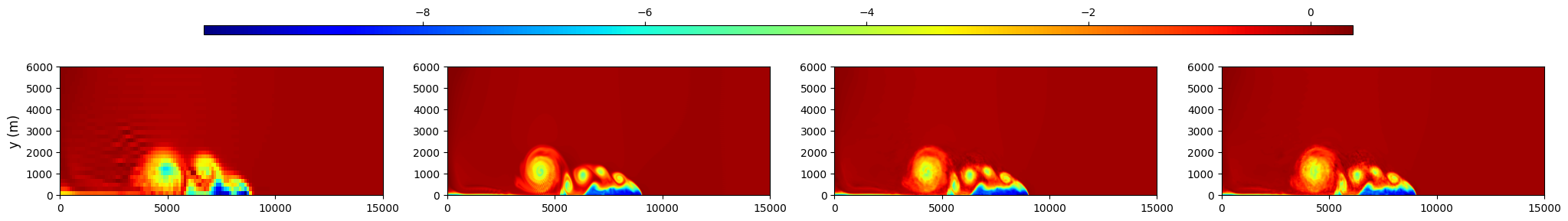}
        \put(12,7){\scriptsize{\textcolor{white}{$t/T=1$}}}
        \put(35,7){\scriptsize{\textcolor{white}{$t/T=1$}}}
        \put(60,7){\scriptsize{\textcolor{white}{$t/T=1$}}}
        \put(85,7){\scriptsize{\textcolor{white}{$t/T=1$}}}
        \put(15,-2){\footnotesize{$I_{LR}$}}
        \put(37,-2){\footnotesize{$I_{HR}$}}
        \put(55,-2){\footnotesize{SR with \emph{m-CNN}}}
        \put(80,-2){\footnotesize{SR with \emph{A-CNN}}}
        \put(88,12){\footnotesize{$\theta'$}}
    \end{overpic}
    
    \caption{Density current, coarse mesh $h = 200$ m: low resolution solution computed by the AV model (first column), reference solution computed by the Smagorinsky model (second column), improvement by the super resolution with \emph{m-CNN} (third column) and \emph{A-CNN} (fourth column) for increasing time $t$ from top to bottom, with $T = 900$ s.}
    \label{fig:200_25_ms_att_comp}
\end{figure}

The high-resolution dataset $I_{HR}$ now consists of 
the solutions provided by the Smagorinsky model with a structured mesh with grid size $h=25$ m and time step $\Delta t = 0.01$ s. 
We sample the simulation $1800$ times
(i.e., once every 5 time steps). We start with 
80\% of the snapshots used for training, while the remaining 20\% are reserved for validation.
The low resolution data $I_{LR}$ are the solutions computed by the AV model 
with two coarser structured meshes with mesh size $h=200, 400$ m and time step $\Delta t = 0.1$ s.
We will consider only the improvements
by super resolution with \emph{A-CNN} 
and \emph{m-CNN} because the computational cost of SR with \emph{Diff} proved to be prohibitive.
Fig.~\ref{fig:400_25_ms_att_comp} and 
\ref{fig:200_25_ms_att_comp} present the comparisons
of $I_{HR}$ and $I_{LR}$ with and without improvements
by the SR.
These comparisons confirm the conclusions drawn from Fig.~\ref{fig:400_100_ms_att_comp}-\ref{fig:200_100_ms_att_comp}, i.e., 
the SR with \emph{m-CNN} outperforms the 
SR with \emph{A-CNN} in terms of visual comparison. 


\begin{figure}[htb!]
    \centering
    \includegraphics[width=.6\linewidth]{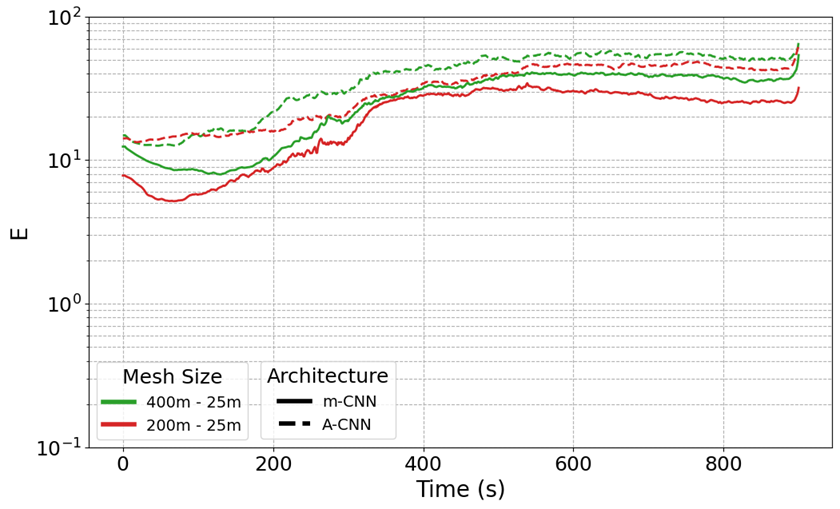}
    \caption{Density current, Smagorinsky model for the reference solution (computed with mesh $h=25$): evolution of error \eqref{eq:L2_norm} for the solutions computed by the AV model with coarse meshes  $h = 400,200$ m 
    enhanced by SR with \emph{A-CNN} and \emph{m-CNN}.
    }
    \label{fig:L2_att_ms}
\end{figure}

For a quantitative comparison of the plots in 
Fig.~\ref{fig:400_25_ms_att_comp}-\ref{fig:200_25_ms_att_comp},
Fig. \ref{fig:L2_att_ms} show the evolution of error (\ref{eq:L2_norm}). 
Like in the case of the rising thermal bubble
benchmark, when the reference solution is 
given by the Smargorinsky model the
errors increase. Compare the curves in Fig.~\ref{fig:L2_norm_S} with the curves in Fig.~\ref{fig:L2_att_ms}. However, unlike the case of the rising thermal bubble
benchmark, the errors are large: over 30\%
for $t > 300$, with the exception of \emph{m-CNN}
with mesh $h=400$ m, which is slightly more accurate.
These larger errors when the reference solution is given by the Smagorinsky model are due to the reduced amount of training data (1800 snapshots vs 9000 used when the reference solution is given by the AV model), an intentional choice for computational efficiency. In fact, 
recall that the mesh for the reference solution by the Smagorinsky (resp., AV) model has size $h = 25$ m (resp., $h = 100$ m). A larger training dataset
could improve the errors in 
Fig.~\ref{fig:L2_att_ms} at the price of an increased computational cost.



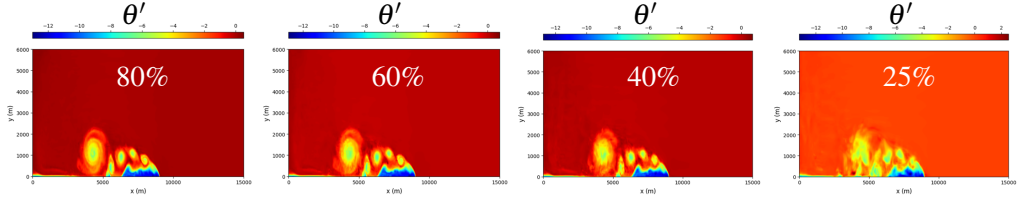
\begin{figure}[htb!]
    \centering
    \setlength{\tabcolsep}{2pt} 
    
    \begin{tabular}{ccc}
        \begin{overpic}[width=0.2\textwidth, grid = false]{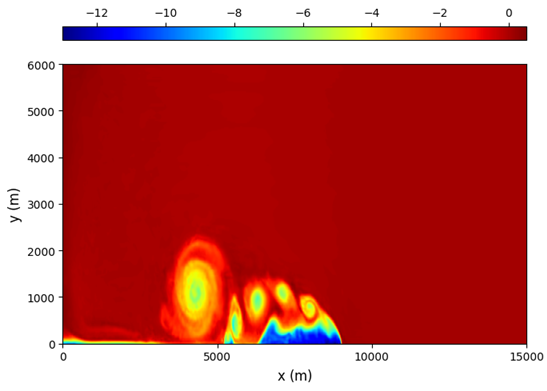}
        \put(45,45){\color{white}{80\%}}
            \put(47,70){\color{black}$\theta '$}
        \end{overpic} 

        \begin{overpic}[width=0.2\textwidth]{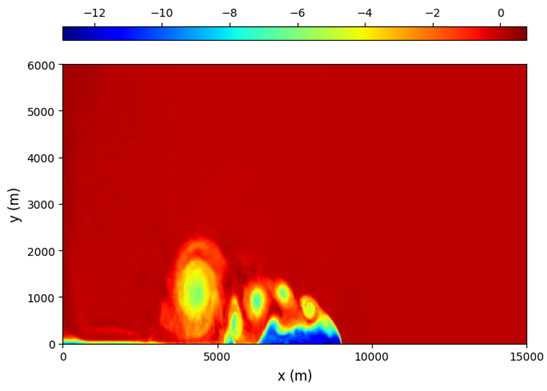}
        \put(45,45){\color{white}{60\%}}
            \put(47,70){\color{black}$\theta '$}
        \end{overpic} 

        \begin{overpic}[width=0.2\textwidth]{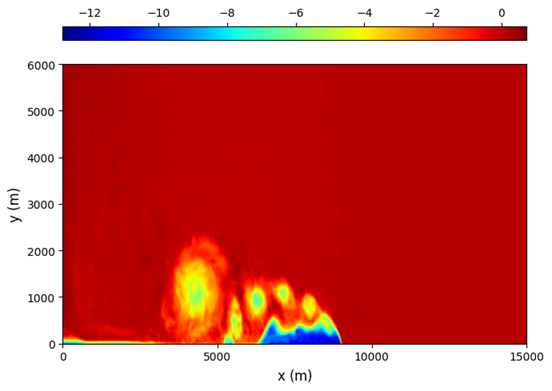}
        \put(45,45){\color{white}{40\%}}
            \put(47,70){\color{black}$\theta '$}
        \end{overpic} 

        \begin{overpic}[width=0.2\textwidth]{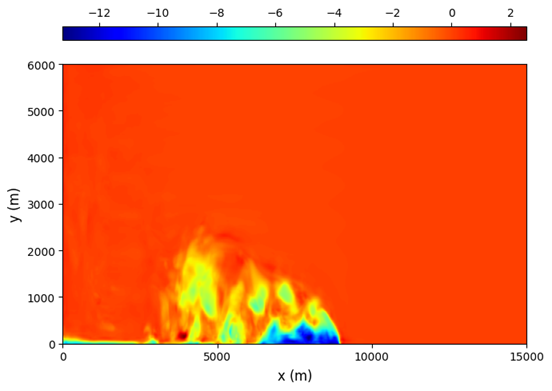}
        \put(45,45){\color{white}{25\%}}
            \put(47,70){\color{black}$\theta '$}
        \end{overpic} \\
    \end{tabular}
    \caption{
    Density current, coarse mesh 
    $h =400$ m: improvements to the low-resolution solution computed by the AV model at time $T = 900$ s by the super resolution with \emph{m-CNN} trained with 
    80\%, 60\%, 40\%, 25\% of the dataset. The $I_{HR}$
    dataset consists of solutions computed by the Smagorinsky model.
    }
    \label{fig:DC_comparison}
\end{figure}

To identify the operational limits of the \emph{m-CNN} architecture, we systematically reduce the training dataset.
Fig.~\ref{fig:DC_comparison} shows the improvements to the low-resolution solution computed by the AV model (mesh $h = 400$ m) at time $T = 900$ by the SR with \emph{m-CNN}
trained with 80\%, 60\%, 40\%, and 25\% of the $I_{HR}$ dataset.
We see that our model demonstrates significant robustness when utilizing 80\% to 60\% of the available data. However, 
when the training uses only 40\% of the dataset, the primary vortex, which is a feature dominated by low-frequency energy scales, begins to show signs of structural degradation. 
In addition, while the general flow physics are captured, 
accuracy is low. The boundaries of model viability are definitely surpassed when the training dataset is restricted to $25\%$.
At this threshold, the architecture is unable to resolve the complex topological characteristics of the flow field. 
The low-frequency vortex structures are severely under-resolved, and high-frequency components exhibit dimensions that are inconsistent with the underlying physics. 

\section{Conclusion} \label{sec:concl}

We considered several super resolution techniques with the goal of improving the accuracy of low-resolution numerical simulations of atmospheric flows. 
The techniques under consideration are a baseline CNN
composed of repeated blocks of bilinear upsampling, convolutional layers, and nonlinear activation functions, 
an attention-enhanced CNN, and a multi-scale version
of the CNN architecture. Additionally, we evaluate a diffusion-based SR model.
The various SR methods were assessed through two standard atmospheric benchmarks: the rising thermal bubble benchmark, i.e., a bubble of warm air that deforms while rising and displays a Rayleigh-Taylor instability with appropriate resolution, and the density current benchmark, which simulates a cold front that propagates
while creating multiple vortices.

We showed that the baseline CNN SR provides an effective framework to learn 
mappings from low-resolution to high-resolution 
images in the case of simpler flows, like in the thermal rising bubble test. This hold true also when the training data is reduced to 60\% of the dataset associated to the entire simulation.
However, the standard CNN-based SR fails to produce
accurate results for more complex flows, like
in the density current test case. Our results
indicated that the attention-enhanced CNN improves accuracy, but not up to a satisfactory level. 
To capture with sufficient accuracy the vortices of different sizes present in the flow, 
a multi-scale \emph{CNN} architecture
with parallel layers and different kernel sizes
was needed. The SR based on multi-scale CNN 
outperformed also a state-of-the-art diffusion model
in terms of accuracy, robustness, and computational 
efficiency. Finally, we showed that a limitation
of the multiscale approach is its sensitivity to the amount of training data. 

\section*{Aknowledgements} \label{sec:thanks}
\textbf{GR} acknowledges the support provided by the European Union - NextGenerationEU, in the framework of the iNEST - Interconnected Nord-Est Innovation Ecosystem (iNEST
ECS00000043 – CUP G93C22000610007) consortium. \textbf{AS} acknowledges the support provided by the European Union - NextGenerationEU (Piano Nazionale di Ripresa E Resilienza (PNRR)) DM 351. \textbf{MG} aknowledges the support provided by Premio Singoli Ricercatori 2025  ``Sviluppo di modelli ridotti per la la fluidodinamica computazionale con  applicazione a campi di moto compressibili'' (PJ\_GEST\_STR\_DIP\_2025\_D262025). We acknowledge the support provided by MUR PRIN “FaReX - Full and Reduced order modelling of coupled systems: focus on non-matching methods and automatic learning” project.

\section*{CRediT Authorship Contribution Statement}

\textbf{Armin Sheidani:} Conceptualization, Methodology, Software, Investigation, Data curation, Formal analysis, Visualization, Writing – original draft. \\
\textbf{Michele Girfoglio:} Conceptualization, Methodology,  Supervision, Validation, Writing – review \& editing. \\
\textbf{Annalisa Quaini:} Methodology, Supervision, Formal analysis, Visualization, Writing – review \& editing. \\
\textbf{Gianluigi Rozza:} Conceptualization, Resources, Supervision, Project administration, Funding acquisition, Writing – review \& editing.
\bibliographystyle{unsrt}
\bibliography{bibliography2}

\end{document}